\documentclass[sigconf,nonacm]{acmart}

\usepackage{graphicx}
\definecolor{Goldenrod}{HTML}{F0E2CE}   
\definecolor{GreenYellow}{HTML}{DCECE9}
\definecolor{pink}{HTML}{FCDCE1}
\usepackage{xcolor}
\usepackage{subcaption}
\usepackage{multirow}
\usepackage{stfloats}
\usepackage{longtable}
\usepackage{booktabs}
\usepackage{array}
\usepackage{tabularx}

\AtBeginDocument{%
  }

\setcopyright{acmlicensed}
\copyrightyear{2025}
\acmYear{2025}
\setcopyright{cc}
\setcctype{by-sa}
\acmConference[DIS '25]{Designing Interactive Systems Conference}{July 5--9, 2025}{Funchal, Portugal}
\acmBooktitle{Designing Interactive Systems Conference (DIS '25), July 5--9, 2025, Funchal, Portugal}\acmDOI{10.1145/3715336.3735753}
\acmISBN{979-8-4007-1485-6/2025/07}

\begin{document}

\title{Exploring the Innovation Opportunities for Pre-trained Models}

\author{Minjung Park}
\email{mpark2@andrew.cmu.edu}
\orcid{0000-0002-8544-4422}
\affiliation{%
  \institution{Carnegie Mellon University}
  \city{Pittsburgh}
  \state{PA}
  \country{USA}
  \postcode{15232}
}
\author{Jodi Forlizzi}
\email{forlizzi@cs.cmu.edu}
\orcid{0000-0002-7161-075X}
\affiliation{%
  \institution{Carnegie Mellon University}
  \city{Pittsburgh}
  \state{PA}
  \country{USA}
}

\author{John Zimmerman}
\email{johnz@cs.cmu.edu}
\orcid{0000-0001-5299-8157}
\affiliation{%
  \institution{Carnegie Mellon University}
  \city{Pittsburgh}
  \state{PA}
  \country{USA}
}

\renewcommand{\shortauthors}{Park, et al.}
\begin{abstract}
Innovators transform the world by understanding where services are successfully meeting customers' needs and then using this knowledge to identify failsafe opportunities for innovation. Pre-trained models have changed the AI innovation landscape, making it faster and easier to create new AI products and services. Understanding where pre-trained models are successful is critical for supporting AI innovation. Unfortunately, the hype cycle surrounding pre-trained models makes it hard to know where AI can really be successful. To address this, we investigated pre-trained model applications developed by HCI researchers as a proxy for commercially successful applications. The research applications demonstrate technical capabilities, address real user needs, and avoid ethical challenges. Using an artifact analysis approach, we categorized capabilities, opportunity domains, data types, and emerging interaction design patterns, uncovering some of the opportunity space for innovation with pre-trained models.  

\end{abstract}



\maketitle

\section{Introduction}
Innovators benefit from understanding what is working, from knowing about the situations where products and services are currently succeeding\cite{kline2010overview, drucker2002discipline}. In many cases, this knowledge helps innovators find a lower risk place to begin product development. By \textit{innovator}, we mean the practitioners who work on new products and services they expect to successfully deploy in the world. These include designers, HCI practitioners, consultants, engineers, business people, executives, administrators, and others working in the commercial and public sectors. Innovators rely on knowledge of what is working, from knowing the situations where products and services are currently co-creating value for customers and service providers. For example, makers of mobile phones noticed the success of click-and-go digital cameras—small cameras that were easy to carry and helped people visually document their lives. They innovated by adding cameras to phones. They made users' lives better by reducing what they needed to carry in their pockets, purses, and bags. Later, innovators noticed the emergent behavior of selfie-taking, and they introduced a forward-facing camera on mobile phones.

In support of knowing what is working, design practitioners often create resources that document successful designs. Online repositories of interaction design patterns provide one example \cite{UIPatterns, UXPin}. These show conventional ways of overcoming frequent interaction challenges. While researchers most often want to know the gaps–to identify what is not being done so they can make a novel contribution–innovators notice and discuss what is working to mitigate risk of creating things people don't want and won't use.

Over the last several years, a growing body of design research explored AI innovation. This research revealed challenges with integrating data science into the enterprise \cite{mao2019data} and challenges HCI/UX practitioners face when trying to envision AI products and services that people want and that can be easily developed \cite{dove2017ux, yang2018mapping, yildirim2023creating}. Researchers noted a very high failure rate for AI initiatives within companies \cite{weiner2022ai}. AI systems failed for technical, financial, user acceptance, and/or ethical issues. Design researchers also noted missing, low-hanging fruit–situations where simple AI could create immediate value but was not developed \cite{yang2016planning, yildirim2022experienced}. They described an \textit{AI innovation gap} in which data science teams envision services customers don't want while HCI/UX teams envision services that cannot be built. To address these problems, researchers developed resources documenting AI capabilities found in commercially successful products and services \cite{yildirim2023creating}, new design processes to help innovators envision better things to build \cite{yildirim2024sketching}, and guidebooks to support prototyping of effective and responsible AI systems \cite{amershi2019guidelines, ibm, googlePAIR, apple}.

The release of ChatGPT in November 2022 spurred huge interest from innovators and a \textquotedblleft gold rush\textquotedblright of investment in creating new AI products and services that make use of pre-trained models. In this paper, we use the term \textit{pre-trained model} to collectively mean Large Language Models, Generative AI, and Foundation Models. We chose this term because the pre-trained aspect of these models offers a major shift for innovators. Instead of collecting data and building a model, innovators can get a faster start by trying models that exist. Pre-trained models offer lots of transfer learning–a model trained to do one thing is also capable of many other tasks. In human terms, transfer learning is like when people learn how to hit a golf ball, they have also developed some of the knowledge and skills needed to hit a baseball, the skill of hitting a thing with a stick. Learning one skill provides learning that works for other things. Pre-trained models initially designed for language translation have unintentionally gained additional skills, such as generating computer code from text descriptions \cite{HUANGg23}. The large amount of transfer learning in pre-trained models makes it unclear what they can and cannot do, let alone how well they might perform different tasks.

While pre-trained models lower the barriers to creating new AI products and services, they also significantly raise costs and challenge business models. For example, some industry analysts estimate that the use of pre-trained models for web search can increase the cost 10 to 100 times \cite{Wangg24}. Moreover, Howell et al. point out that running large models can be expensive and scale poorly with increased usage \cite{howell2023economic}. Question-answering with pre-trained models complicates the web search business model. Traditionally, web search service providers get paid when users click on ad links. When web search questions are asked to pre-trained models, the user sees responses in the form of answers to their question, not a list of ads that will generate revenue for web search providers. For these reasons, it is difficult for innovators to know what pre-trained models can or cannot do, where they might produce more value than costs, where users might willingly accept and use new innovations, and where these innovations do not introduce ethical challenges or unintended harms that significantly diminishes a system's overall value. Innovators lack a resource that tells them where pre-trained models are currently succeeding.

We wanted to help innovators make better choices about what to make using pre-trained models. We wanted to develop a resource documenting situations where pre-trained models are likely to succeed. Unfortunately, commercial applications don't work well for building this resource. While they capture the technical aspects of a capability, current examples don't provide evidence of financial viability, user acceptance, or avoidance of ethical harms. Pre-trained models are caught up in a \textit{hype cycle}  \cite{dedehayir2016hype, Gartner}. They are largely funded by venture capital, and it is not clear which application areas will prove to be popular and financially viable. Venture capital-supported companies often engage in what might be classified as \textit{predatory pricing} \cite{areeda1975predatory}; they make their services available to customers at prices well below actual costs. They are rushing to develop customers and build market share before raising their prices to a level that can cover their costs. Today, even the most popular pre-trained model services, like Microsoft's Co-Pilot for programmers, cost significantly more to operate than they charge customers for the service \cite{dotan2023big, TheRegister}.

As the next best option, we chose to analyze the growing number of pre-trained model applications developed by HCI researchers. These applications demonstrate a technical capability, and they also address real user needs. In addition, with the growing HCI community interest in responsible AI, researchers developing these applications most often consider the possible ethical harms. The one risk of failure that the research applications almost never address is financial viability. The systems developed by HCI researchers rarely discuss or demonstrate that their application can generate more value than costs. While imperfect, we felt a resource documenting what pre-trained models can do, that  users want, and that avoids ethical concerns would be better than the current state (better than nothing). We analyzed a corpus of 85 HCI research applications, categorizing their domains, the technical capabilities, the minimum model performance needed to create value for users, and emerging design patterns for human-AI interaction.

This paper makes three contributions. First, it provides the first draft of a resource showing where innovation with pre-trained models might lead to success. This can help innovators make better choices for where to start. Second, it provides a high-level perspective on the kinds of applications HCI researchers are exploring, revealing gaps for new research. Third, it offers some emerging interaction design patterns that can help both innovators and HCI researchers as they create applications, products, and services that make use of  pre-trained models.

\section{Related Work}

We draw on four areas of HCI research: research on HCI/UX practitioners' struggles with AI, development and assessment of AI guidelines,  research to help with ideation and project selection, and emerging work on innovation with pre-trained models.

\subsection{HCI Investigations of Practitioners}
HCI has investigated AI innovators, particularly the challenges of data scientists trying to collaborate with other stakeholders within the enterprise \cite{kross2021orienting, mao2019data, lam2023model, yang2019sketching, yildirim2023investigating, dove2017ux, yang2020re}. Studies documented the struggles data scientists face in effectively communicating what they do and what they mean when talking about a model's likely performance \cite{mao2019data}. Research shows AI innovators often choose projects that are too technically challenging \cite{yang2020re}. Research suggests that data scientists envision AI services customers do not want while UX/HCI teams envision services that cannot be built \cite{kross2021orienting, lam2023model, yang2019sketching, yildirim2023investigating, dove2017ux, yang2020re}.

Research noted that UX/HCI practitioners often struggle to understand AI capabilities–what AI can and cannot do \cite{dove2017ux, yang2018mapping}. UX/HCI practitioners who have successfully integrated AI into their innovation processes have internalized abstractions AI capabilities \cite{yildirim2022experienced}. They use examples of specific capabilities to communicate to other design/HCI practitioners and to data science collaborators. Yildirim et al. found that  co-locating UX/HCI practitioners and data scientists might improve collaboration and innovation \cite{yildirim2022experienced}. Close collaboration fosters better communication, enabling more effective and practical AI innovation. Research captures how some UX/HCI practitioners have begun to develop resources and frameworks that document AI capabilities to improve their ability to envision things that can be built \cite{yildirim2023creating}.

\subsection{Resources and Methods for Improving Ideation}
HCI research has explored how to improve ideation of AI concepts that are technically feasible and desired by users \cite{liu2024human, yildirim2023creating, kocielnik2019will}. This addresses two of the four main reasons AI projects fail. One approach specifically focuses on the use of simple AI and on discovery of situations where moderate, model performance creates customer value \cite{yildirim2023creating}. This addresses the observation that innovators are overlooking the \textquotedblleft low-hanging fruit.\textquotedblright These researchers assembled a corpus of 40 commercially successful AI features covering 14 industrial domains. Interestingly, 25 of these 40 features required only moderate model performance to generate customer value \cite{yildirim2023creating}.

To better understand the impact of model performance, HCI researchers developed a task expertise-model performance matrix. They referred to this as the opportunity space for AI innovation \cite{yildirim2024sketching, gmeiner2023dimensions}. This conceptual model aids interdisciplinary teams in exploring AI concepts that are both valuable and easy to develop, promoting a focus on low-risk, high-value applications. Collectively, this work has improved the ideation of AI concept; however, the work almost entirely focuses on narrow AI.

\subsection{Innovation with Pre-trained Models}
The 2022 release of ChatGPT \cite{openai} triggered a wave of interest and concern around pre-trained models. Unlike narrow AI, which is designed to perform a single specific task, pre-trained models leverage transfer learning to provide many unintended tasks \cite{qiu2020pre, weiss2016survey}. This raises new challenges for understanding what these models might be able to do. The use of pre-trained models also brings significant benefits by lowering the effort required for development \cite{han2021pre}. Since there is no longer a need to collect extensive data and build a model from scratch, innovators can simply use a pre-trained model to test their ideas. However, operating pre-trained models comes with much higher costs \cite{Smith23}. For example, using a pre-trained model for web search can cost 10 to 100 times more than a traditional web search. Additionally, these models are prone to \textquotedblleft hallucinations," where they generate incorrect or nonsensical outputs. 

HCI research has begun to explore the challenges of developing systems using pre-trained models. For instance, researchers found that when you fix a model's error, this can cause an error that was previously fixed to reappear \cite{zamfirescu2023herding}. HCI researchers are also generating many new applications that demonstrate the capabilities of pretrained models and illustrate how they might create value for different kinds of users. HCI researchers also conducted a systematic review to explore how the HCI community perceives the use of LLMs. They performed a systematic literature review of CHI papers and identified domains, roles in HCI projects and key concerns \cite{pang2025understanding}.

Our research builds on prior efforts to help innovators envision what they can create with AI. While earlier studies have primarily focused on narrow AI, we shift the focus to pre-trained models, aiming to advance available resources and expand the scope of possibilities for innovation.

\section{Method}
We wanted to help innovators by identifying what pre-trained models can do that can create value in the world. We took a designerly approach, playing with the pre-trained models as a way of understanding what they might or might not be able to do. We thought of this as engaging with pre-trained models as a design material, building on prior HCI work that discusses AI as a design material \cite{dove2017ux, yildirim2022experienced, feng2023ux, feng2024canvil, moore2023failurenotes, liao2023designerly}. We planned to design things to gain a felt understanding of what is possible.

We began by identifying a number of tasks where we assumed pre-trained models would work, and then developed prototypes that demonstrate the capability.  We explored many tasks, including providing feedback on posters, analyzing tabular information, classifying messages, scanning resumes, and standardizing formats for references and citations. These efforts all failed due to technical, ethical, and user acceptance issues. For instance, when we asked to filter and analyze specific information from  tabular data, it faced technical and ethical limitations, often missing critical information or producing biased results. Moreover, when it provided feedback on posters, we questioned whether users would accept the quality of that feedback. For almost all of the applications we tried, we could not achieve an acceptable level of performance. Our pre-trained model applications just created more work for people, not less. We found this process frustrating. Adding to the frustration was a general level of uncertainty surrounding pre-trained models. When we could not get systems to do what we wanted, we could not easily tell if the problem was our prompting skills or if we were simply asking too much of the pre-trained models. Our frustration in ideating buildable ideas seemed eerily similar to Yang et al.'s work on Sketching NLP \cite{yang2019sketching}, where researchers struggled to envision useful ideas that could be built.

Our frustration drove us to consider a different goal and a different approach. We reframed the problem. Instead of asking what pre-trained models can do, we shifted to asking, \textquotedblleft what have people been successful at getting pre-trained models to do?\textquotedblright Instead of playing with and building things using pre-trained models, we shifted to analyzing applications that made effective use of this technology.

\subsection{Selecting a Corpus}

Inspired  by the success of Yildirim et al.'s use of commercially successful AI features to build a taxonomy of AI capabilities, we wanted to follow a similar approach with a focus on pre-trained models. We wanted to help innovators avoid the four main causes of AI project failure \cite{weiner2022ai}: (i) cannot achieve the minimally acceptable model performance, (ii) development and operational costs outweigh the application's value, (iii) users will not accept and use the application (often because the system does not address a real user need), and (iv) the application has ethical challenges that create unintended harm. However, we struggled to find a corpus of successful applications. 

\begin{figure}[h]
  \centering
  \includegraphics[width=1.0\linewidth]{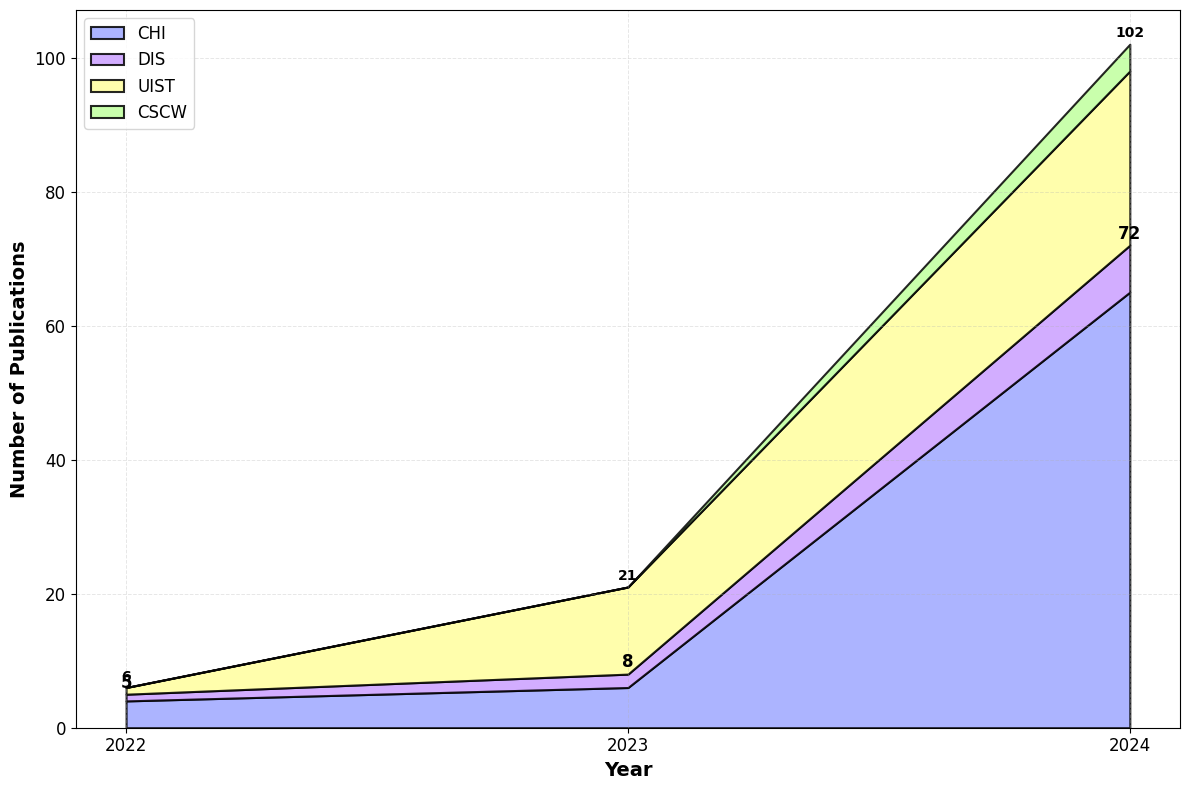}
  \caption{The number of papers presenting applications made with pre-trained models over the past three years from CHI, DIS, UIST and CSCW. We start in 2022 as this is when Open AI released ChatGPT and kicked off the public's interest in pre-trained models.}
  \label{fig:corpus}
\end{figure}

We chose not to document commercial applications due to the current hype cycle. We worried that current commercial applications—funded by venture capital and using \textquotedblleft predatory pricing\textquotedblright— could mislead innovators into inferring impossible financial models for inferring user needs that do not exist. We reasoned that research HCI applications could serve as a more valuable proxy for commercial success. We explored applications from four HCI venues: CHI, DIS, CSCW, and UIST. The number of papers presenting applications made with pre-trained models over the past three years from these four venues has grown rapidly (shown in Figure\ref{fig:corpus}).

We chose to exclude UIST and CSCW. UIST primarily focuses on exploring speculative technologies with less attention to user needs,  CSCW is more oriented toward theories of human behavior and the impact of AI on human-to-human and human-to-AI collaboration than in demonstrating technology users might find valuable. We chose to exclusively focus on CHI and DIS for their alignment with our research goals of technical feasibility, user acceptance, and avoidance of ethical concerns.

\begin{figure}[h]
  \centering
  \includegraphics[width=1.0\linewidth]{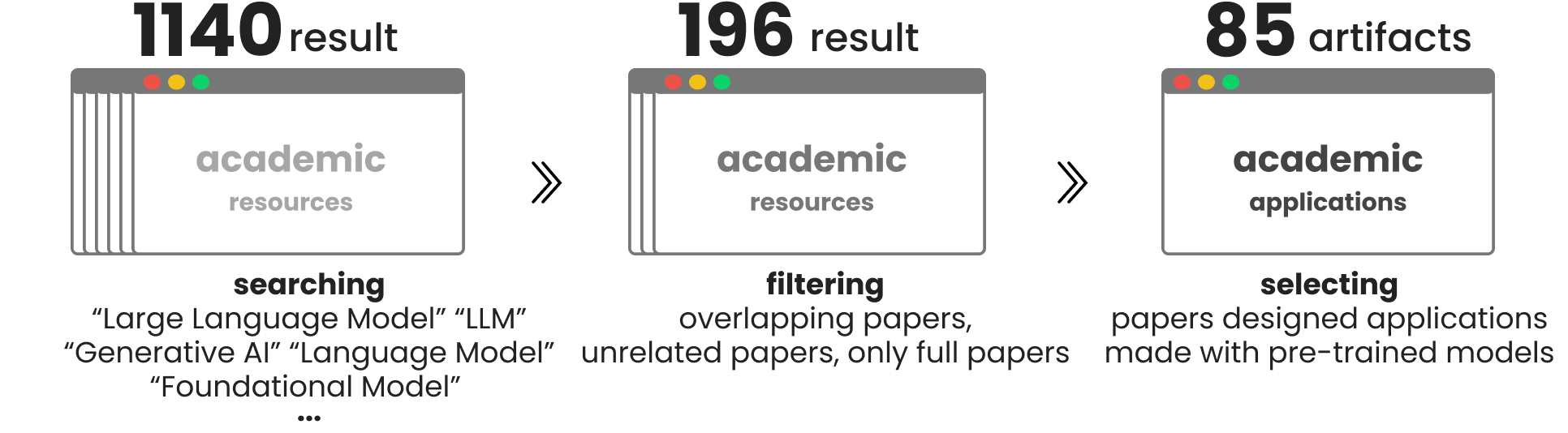}
  \caption{Illustrating the search, filtering, inclusion, and exclusion process.}
  \label{fig:corpus2}
\end{figure}

Notably, we were interested in the applications built with pre-trained models, not in the research questions being asked. Our focus was on making a valuable resource that innovators might use as a starting place for envisioning new AI products and services that make use of pre-trained models. To create the collection, we searched in the ACM SIGCHI database and its mirror (\url{dl.acm.org} and \url{programs.sigchi.org}). We used the terms \textquotedblleft Large Language Model\textquotedblright, \textquotedblleft LLM\textquotedblright, \textquotedblleft Foundation Model\textquotedblright, \textquotedblleft Language Model\textquotedblright, \textquotedblleft Large Language\textquotedblright and \textquotedblleft Generative AI\textquotedblright. Our search returned 1140 publications (Large Language Model: 212; LLM: 240;  Large Language: 221; Generative AI: 189, Language Model: 264; Foundation Model: 14)  in their title or abstract. Initially, we focused on overlapping papers that repeatedly appeared in searches using multiple search terms. For example, a title like \textquotedblleft Challenges and Opportunities in Designing with Generative AI and Large Language Models for HCI (hypothetical example)\textquotedblright could be found using both \textquotedblleft Generative AI\textquotedblright and \textquotedblleft Large Language Model\textquotedblright as search terms. Then, we filtered out papers that mentioned the search terms in the abstract but were unrelated to them in the main content of the paper. Finally, we only considered full papers, narrowing the dataset to 196. From these, we identified 85 full papers that designed applications with pre-trained models (see Appendix \ref{tab:artifactlist}  for the full list and the detailed flow of inclusion is shown in Figure\ref{fig:corpus2}).

\subsection{Artifact Analysis}
We used artifact analysis \cite{hanington2019universal, philip23, janlert2017things, use15} to guide our analysis of  the 85 AI applications (see Appendix\ref{tab: artifactlist} for the full list). Artifact analysis comes from social anthropology. Researchers analyze the things people use as a way of better understanding people. HCI researchers have used this method to explore highly uncertain tasks like the design of a robot. DiSalvo et al. developed an initial understanding of which features and dimensions of a humanoid robot's face most significantly influence people's perception of its humanness \cite{disalvo2002all}. They used artifact analysis to gain insights into which specific facial features of robots are effective in evoking anthropomorphism. Similarly, Odom et al. applied artifact analysis in the context of slow design cases with the ultimate goal of producing new concepts that could support innovative practices within an expanded, design-oriented theoretical framework \cite{odom2022extending}. 

Given our focus on trying to guide innovators toward failsafe opportunities for innovation, we chose four application aspects to investigate. We looked at the domain (e.g., healthcare, hospitality). We extracted pre-trained model capabilities–what pre-trained models can reasonably do. We inferred an application's task expertise (how hard is the task for a person) and model performance (the minimal performance the AI needs to create value for the user). Finally, we searched  for emerging design patterns showing how users might effectively engage with applications that use pre-trained models.

\subsubsection{Identify Domains}
Yilidirim et al.'s work on AI capabilities covered 14 industrial domains. We used this structure as a starting place.  To infer the domain, we looked at how authors described the intended users of their system. For example, a system made for doctors would get placed in healthcare. In cases where the user did not map into one of the 14 industrial domains, we added a new domains to the list.

\subsubsection{Extract Capabilities}
We extended the process used by prior literature \cite{yildirim2023creating}, including their focus on capturing AI capabilities (what AI can do) and not mechanism (how it makes an inference). We followed their bottom up, inductive approach. We first detailed specific capabilities for each application. This resulted in 294 specific capabilities. We then collapsed these into relevant clusters, hiding unnecessary detail and keeping the focus on what a pre-trained model could reasonably do.

As part of this bottom up process, we slowly evolved a grammar for describing a capability: [Capability (action verb)] + [Output form or structure] + [Input data]. For example: a system that produced a transcript from audio would appear as [Transcribe] + [into Text] + [from Speech]. This structure allowed us to capture each capability as a sentence and then compare it to the other capabilities. We created consistent language across the examples. For example, an individual capability might use the term Talk, Speak, or Vocalize; however, when looking across the set of capabilities, we would choose the best term to bring these capabilities together, in this case, Vocalize.

To guide this inductive process, we would individually document capabilities and then meet as a group to discuss and reach a consensus on the grammar and on the terms. Throughout this process, we kept a tight focus on making choices that would make this resource of capabilities useful to innovators. Similar to Yildirim \cite{yildirim2023creating}, this meant we needed to constantly consider the relevant level of granularity, the generality of the specific words we chose, and the breadth that our capabilities conveyed. We worried that innovators might incorrectly assume that the pre-trained models were more capable than our dataset really indicated. For example, one capability details how pretrained models can answer a question about a product when given the description of a product. We intentionally kept the term \textquotedblleft product\textquotedblright for this example to avoid innovators thinking a pre-trained model could answer a question about anything that could be described.

\subsubsection{Infer Expertise and Performance}
Yildirim et al.'s work showed that when designers were given a set of commercially successful AI capabilities, they still envisioned things that could not be built \cite{yildirim2023creating}. To overcome this challenge, they created the task-expertise/model-performance matrix. Taking their initial failure as a lesson, we followed their process and mapped the 85 applications in terms of how hard a task was for a human to perform (three levels: expertise, typical adult, less than a typical adult), and in terms of the minimal level of model performance needed to create user value (moderate, good, and excellent). As an example, in situations where a user wants to find an image of a cat from an image dataset, the system would need only moderate model performance. If the user needed to find all the images within the dataset that showed a cat, then the system would need excellent model performance. Based on the application descriptions from the research papers and by leveraging Yildirim et al's process, we made collective, subjective inferences for task-expertise and for model-performance.

\subsubsection{Explore Interaction Design Patterns}
The developers of pre-trained models often allow people to interact with the model using a prompt-based interface like the one used by ChatGPT. All of the research applications placed an Human-AI interface between users and the pre-trained model's prompt interface. We analyzed these interfaces to discover emerging design patterns \cite{borchers2000pattern, norman1986user}. We looked at the interface images as well as text describing the human-AI interaction. We followed a process similar to Yang et al. \cite{yang2016planning}, in their work to discover design patterns for adaptive UIs used in mobile apps. We first identified the \textquotedblleft problem\textquotedblright that the pattern addressed. This could be the user's problem or the service provider's problem.  In most cases, this was not explicitly stated. Next, we looked for commonalities in the interaction flow and sequencing, and in the layout of the elements and features of the different interfaces. This process revealed seven emerging design patterns. For each pattern, we noted the problem and how the interface addresses the problem. In the findings section, we describe each pattern along with examples from our corpus of research applications.

\section{Findings}
The findings of our artifact analysis revealed the domains where pre-trained models create value, the capabilities researchers employed in their applications, the types of input and output data used, and some emerging design patterns for interacting with applications that employ pre-trained models.

\subsection{Domains Where Pre-trained Models Create Value}
\begin{figure}[h]
  \centering
  \includegraphics[width=1.0\linewidth]{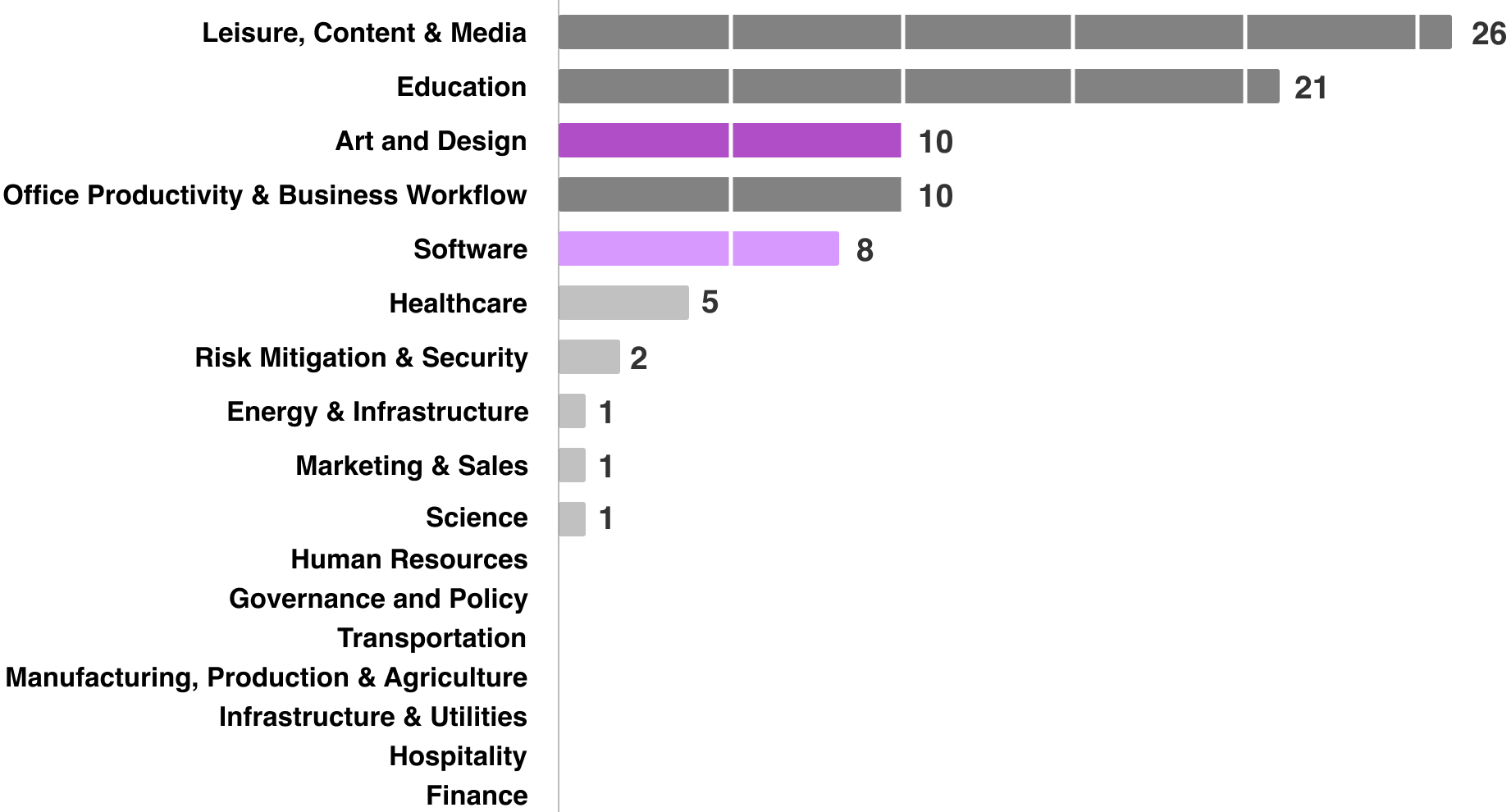}
  \caption{ Mapping of pretrained model applications to industrial domains published in Yildirim et al\cite{yildirim2023creating} where AI has traditionally created value.}
  \label{fig:domaintable}
\end{figure}

Many of the applications in our corpus focused on two domains, leisure (26) and education (21) (see Fig. \ref{fig:domaintable}). We discovered that other applications in our corpus focused on  office productivity (10), healthcare (5), security (2), energy (1), marketing (1), and science (1). Interestingly, none of the applications were related to finance, government \& policy, hospitality, human resources, manufacturing \& agriculture, or transportation (domains from \cite{yildirim2023creating}). Eighteen applications did not fit Yildirim et al.'s  domains. These included creativity tools for professional programmers, and tools to help professional designers or other creatives. We cateogrized these into two domains: \textit{Art \& Design} and \textit{Software} , indicated in the figure below(Fig. \ref{fig:domaintable}) in purples.  

\subsection{Pre-trained Model Capabilities}
\setlength{\arrayrulewidth}{0.5mm} 
\setlength{\tabcolsep}{13pt} 
\renewcommand{\arraystretch}{1.5} 

\newcolumntype{L}[1]{>{\raggedright\arraybackslash}p{#1}}

\begin{table*}[]
\caption{\textbf{Pre-trained Model Capabilities}: (Left to Right) 33 capability clusters, 13 specific actions, and three high-level capability themes. (LEGEND: \colorbox{pink}{[Capability (action verb)]} + \colorbox{GreenYellow}{[Output form or structure]} + \colorbox{Goldenrod}{[Input data]} )}

\resizebox{\textwidth}{!}{%
\begin{tabular}{p{15cm} |p{8cm}| p{1cm}}

\cline{1-3}
\textbf{\colorbox{pink}{Capability as {[}Action verb{]}} + \colorbox{GreenYellow}{{[}Ourtput Form{]}} + \colorbox{Goldenrod}{{[}Input Data{]}}}              & \textbf{capability actions and definition}                                                                                                                & \multicolumn{1}{l}{\textbf{capability themes}}                                                     \\ \cline{1-3}
\colorbox{pink}{Render} an \colorbox{GreenYellow}{image} based on a \colorbox{Goldenrod}{topic, mood, tone, keywords, or description} (21)                  & \multirow{3}{*}{\begin{tabular}[c]{@{}l@{}}\textbf{Render(34)}\\ Generate a desired image.\end{tabular}}                                                           & \multirow{7}{*}{\textbf{\begin{tabular}[c]{@{}c@{}}Generate\\ New\\ Content\\ (66)\end{tabular}}}  \\
\colorbox{pink}{Render} a \colorbox{GreenYellow}{persona image} based on a \colorbox{Goldenrod}{persona description} (2)                                    &                                                                                                                                                           &                                                                                                  \\
\colorbox{pink}{Render} an \colorbox{GreenYellow}{image that communicates a tone or mood} based on an \colorbox{Goldenrod}{image} (11)                       &                                                                                                                                                           &                                                                                                  \\ \cline{1-2}
\colorbox{pink}{Write} a \colorbox{GreenYellow}{description} based on \colorbox{Goldenrod}{image} (10)                                                       & \multirow{4}{*}{\begin{tabular}[c]{@{}l@{}}\textbf{Write(32)}\\ Generate a specific form of text, \\[-0.5em] like a story, dialog, description, or questions.\end{tabular}} &                                                                                                 \\
\colorbox{pink}{Write} a \colorbox{GreenYellow}{story} based on a \colorbox{Goldenrod}{topic or description} (7)                                            &                                                                                                                                                           &                                                                                               \\
\colorbox{pink}{Write} a \colorbox{GreenYellow}{description} based on \colorbox{Goldenrod}{keywords} (9)                                                    &                                                                                                                                                           &                                                                                                  \\
\colorbox{pink}{Write} a \colorbox{GreenYellow}{character's response} based on \colorbox{Goldenrod}{dialog} (6)                                            &                                                                                                                                                           &                                                                                             \\ \cline{1-3}
\colorbox{pink}{Code} into \colorbox{GreenYellow}{computer code} based on a \colorbox{Goldenrod}{task description} (11)                                      & \begin{tabular}[c]{@{}l@{}}\textbf{Code (11)}\\ Transform a description into a computer program.\end{tabular}                                                       & \multirow{4}{*}{\textbf{\begin{tabular}[c]{@{}c@{}}Transform\\ Content \\(24)\end{tabular}}}       \\ \cline{1-2}
\colorbox{pink}{Transcribe} into \colorbox{GreenYellow}{text} from \colorbox{Goldenrod}{speech} (8)                                                         & \begin{tabular}[c]{@{}l@{}}\textbf{Transcribe (8)}\\ Transform speech into text.\end{tabular}                                                                      &                                                                                                 \\ \cline{1-2}
\colorbox{pink}{Translate} into a \colorbox{GreenYellow}{description} from \colorbox{Goldenrod}{computational code} (3)                                     & \begin{tabular}[c]{@{}l@{}}\textbf{Translate (3)}\\ Transform a computer program into a description.\end{tabular}                                                  &                                                                                                 \\ \cline{1-2}
\colorbox{pink}{Vocalize} into \colorbox{GreenYellow}{speech} based on a \colorbox{Goldenrod}{transcript} (2)                                               & \begin{tabular}[c]{@{}l@{}}\textbf{Vocalize (2)}\\ Transform text into speech.\end{tabular}                                                                        &                                                                                                  \\ \cline{1-3}
\colorbox{pink}{Answer} a \colorbox{GreenYellow}{question about a product} based on a \colorbox{Goldenrod}{product description} (2)                        & \multirow{3}{*}{\begin{tabular}[c]{@{}l@{}}\textbf{Answer (28)}\\ Understand content and questions \\[-0.5em]to provide answers.\end{tabular}}                             & \multirow{22}{*}{\textbf{\begin{tabular}[c]{@{}c@{}}Understand\\ Content\\(204)\end{tabular}}}      \\
\colorbox{pink}{Answer} a \colorbox{GreenYellow}{question informed by context} based on \colorbox{Goldenrod}{what was mentioned earlier} (17)              &                                                                                                                                                           &                                                                                                 \\
\colorbox{pink}{Answer} \colorbox{GreenYellow}{how to do something} based on a \colorbox{Goldenrod}{question about programming or} \colorbox{Goldenrod}{scientific knowledge} (9) &                                                                                                                                                           &                                                                                                   \\ \cline{1-2}
\colorbox{pink}{Rank} \colorbox{GreenYellow}{program function} based on a \colorbox{Goldenrod}{point in computer code} (2)                                  & \multirow{4}{*}{\begin{tabular}[c]{@{}l@{}}\textbf{Rank (8)}\\ Understand and order actions, elements,\\[-0.5em] and qualities.\end{tabular}}                                &                                                                                                  \\
\colorbox{pink}{Rank} \colorbox{GreenYellow}{personas} based on \colorbox{Goldenrod}{description} (1)                                                       &                                                                                                                                                           &                                                                                              \\
\colorbox{pink}{Rank} \colorbox{GreenYellow}{color scheme} based on \colorbox{Goldenrod}{description} (1)                                                   &                                                                                                                                                           &                                                                                               \\
\colorbox{pink}{Rank} \colorbox{GreenYellow}{keyword suggestions} based on a \colorbox{Goldenrod}{point in prompt} (4)                                      &                                                                                                                                                           &                                                                                                \\ \cline{1-2}
\colorbox{pink}{Find similar} \colorbox{GreenYellow}{keywords or document} based on \colorbox{Goldenrod}{dialog, documents, or description} (24)            & \multirow{2}{*}{\begin{tabular}[c]{@{}l@{}}\textbf{Find Similar (28)}\\ Find similar content.\end{tabular}}                                                        &                                                                                                  \\
\colorbox{pink}{Find similar} \colorbox{GreenYellow}{element in the image} based on a \colorbox{Goldenrod}{group of images} (4)                             &                                                                                                                                                           &  \\ \cline{1-2}
\colorbox{pink}{Identify} an \colorbox{GreenYellow}{inappropriate or offensive response} based on \colorbox{Goldenrod}{dialog} (3)                          & \multirow{5}{*}{\begin{tabular}[c]{@{}l@{}}\textbf{Identify (27)}\\ Recognize specific things in content.\end{tabular}}                                            &                                                                                                  \\
\colorbox{pink}{Identify} \colorbox{GreenYellow}{if a person is in an image} from a \colorbox{Goldenrod}{tagged images} (2)                                 &                                                                                                                                                           &                                                                                                 \\
\colorbox{pink}{Identify} \colorbox{GreenYellow}{sections or elements (problems, methods)} based on \colorbox{Goldenrod}{research paper or dialog} (9)      &                                                                                                                                                           &                                                                                                  \\
\colorbox{pink}{Identify} \colorbox{GreenYellow}{the argument} from a \colorbox{Goldenrod}{document, image, or dialog} (3)                                  &                                                                                                                                                           &                                                                                           \\
\colorbox{pink}{Identify} \colorbox{GreenYellow}{the sentiment} from a \colorbox{Goldenrod}{document, image, or dialog} (10)                                &                                                                                                                                                           &  \\ \cline{1-2}
\colorbox{pink}{Interpret} \colorbox{GreenYellow}{explanation} based on professional terms (7)                                        & \multirow{3}{*}{\begin{tabular}[c]{@{}l@{}}\textbf{Interpret (21)}\\ Understand the subtext,\\[-0.5em] the meaning of the content.\end{tabular}}                             &                                                                                                \\
\colorbox{pink}{Interpret} a \colorbox{GreenYellow}{question to ask someone} based on \colorbox{Goldenrod}{dialog or story} (12)                            &                                                                                                                                                           &  \\
\colorbox{pink}{Interpret} \colorbox{GreenYellow}{reason} based on a \colorbox{Goldenrod}{professional knowledge} (2)                                       &                                                                                                                                                           &  \\ \cline{1-2}
\colorbox{pink}{Summarize} into \colorbox{GreenYellow}{keywords or bullet list} based on a \colorbox{Goldenrod}{document, dialog, or diary} (35)            & \multirow{2}{*}{\begin{tabular}[c]{@{}l@{}}\textbf{Summarize (58)}\\ Summarize content.\end{tabular}}                                                              &  \\
\colorbox{pink}{Summarize} into a \colorbox{GreenYellow}{few sentences} based on \colorbox{Goldenrod}{documents, stories, or dialog} (23)                   &                                                                                                                                                           &  \\ \cline{1-2}
\colorbox{pink}{Refine} \colorbox{GreenYellow}{docment's tone of voice} based on \colorbox{Goldenrod}{document and tone request} (23)                       & \multirow{3}{*}{\begin{tabular}[c]{@{}l@{}}\textbf{Refine (34)}\\ Improve the quality of the content.\end{tabular}}                                               &  \\
\colorbox{pink}{Refine} \colorbox{GreenYellow}{fix grammar error} based on \colorbox{Goldenrod}{text} (2)                                                   &                                                                                                                                                           &  \\
\colorbox{pink}{Refine} \colorbox{GreenYellow}{into a more explicit and effective prompt} based on a \colorbox{Goldenrod}{vague prompt} (9)                 &                                                                                                                                                           &  \\ \cline{1-3}
\end{tabular}%
}
\label{tab: capability}
\end{table*}\label{tab: capability}
We inferred 294 individual capabilities (see Appendix\ref{tab: 241capabilitytable} for the full list). These included things like \textit{\colorbox{pink}{summarize} a \colorbox{GreenYellow}{web page} based on \colorbox{Goldenrod}{title} \colorbox{Goldenrod}{and}\colorbox{Goldenrod}{initial}\colorbox{Goldenrod}{paragraphs}}, \textit{\colorbox{pink}{transform} into a \colorbox{GreenYellow}{JSON file} from a \colorbox{Goldenrod}{pdf}}, and \textit{\colorbox{pink}{identify} \colorbox{GreenYellow}{key features} based on \colorbox{Goldenrod}{children's} \colorbox{Goldenrod}{doodles}} 

\bigskip
\textbf{LEGEND:} \par
\colorbox{pink}{[Capability]} + 
\colorbox{GreenYellow}{[Output form or structure]} + 
\colorbox{Goldenrod}{[Input data]}
\bigskip

We clustered these capabilities based on overlaps. This resulted in 33 capabilities (Table \ref{tab: capability}, column 1). Clustering organized capabilities by their actions and the kinds of data they took as input or produced as output. The 33 capabilities clustered into 13 specific actions (Table \ref{tab: capability}, column 2). 

We further clustered the 13 actions into three high-level categories (Table \ref{tab: capability}, column 3). This was based on the quantity and form of input and output data. \textit{Generate New Content} represents situations where pre-trained models take in a small amount of data describing what users want and return a large amount of content. \textit{Transform Content} represents situations where pre-trained models take in and return approximately equal amounts of content, and where that content gets returned in a new form. \textit{Understand Content} represents situations where pre-trained models take in lots of content and return small amounts of content that characterizes the content the user provided. More than half of capabilities, 69.4\% (204 of 294) fit understand content, 22.4\% (66 of 294) fit generate new content, and 8.2\% (24 of 294) fit transform content. Among the 13 specific capabilities, 7 fit understand content, more than half of the total. Two capabilities fit generate new content, and four fit transform content. We observed a tendency for the main capability of an application to shift from \textit{generate new content} to \textit{understand content} over the past three years of research applications (papers) we analyzed.

The 13 specific capabilities included in each category are as follows:
\begin{itemize}
    \item \textit{\textbf{Generate New Content (66)}}: render (34) and write (32)
    \item \textit{\textbf{Transform Content (24)}}: Code (11), transcribe (8), translate (3), and vocalize (2)
    \item \textbf{\textit{Understand Content (204)}}: summarize (58), refine (34), answer (28), find similar (28), identify (27), interpret (21), and rank (8)
\end{itemize}

Several of the 13 capabilities were more frequently utilized in the applications. For instance, summarize (58), render (34), and refine (34) were the most used. Summarize and refine fit \textit{Understand Content}, while render fit \textit{Generate New Content}. Within \textit{Generate New Content}, both capabilities write and render had over 30 occurrences. Similarly, in \textit{Understand Content}, six out of the seven specific capabilities had over 20 occurrences: summarize (58), refine (34), answer (28), find similar (28), identify (27), and interpret (21). Interestingly, the highest count in the \textit{Transform Content} category was code, with 11 occurrences. The four capabilities with the lowest counts - code(11), transcribe(8), translate(3), and vocalize(2) - all fit \textit{Transform Content}.

\subsubsection{Input/output Data Types}
89\% (261) of the applications used text as input, and 85\% (251) produced text as output. This included various forms of text including stories, descriptions, research papers, dialogues, computer code, and lists of keywords. 8\% (25) used image data as input, and 13\% (39) produced images as output. Input and output images included people, objects, conceptual examples, sketches, interiors, and color palettes. They also included complete images and segments and elements extracted from images. 2.7\% (8) used audio for input, and 1\% (3) output audio. In all cases, audio was of human speech. Depth maps were used as output by one application (0.34\%). None of the applications appeared to use time series data (e.g., web usage logs, medical records), graphs/network data (e.g., relationships, clusters), or sensor data (e.g., motion, non-vocal sound, humidity, radar) as input or output, even though these are frequently used for narrow AI systems.

\subsection{Task-expertise and Model-performance}

\begin{figure*}[t]
  \centering
  \begin{subfigure}{0.47\textwidth}
    \centering
    \includegraphics[width=\linewidth]{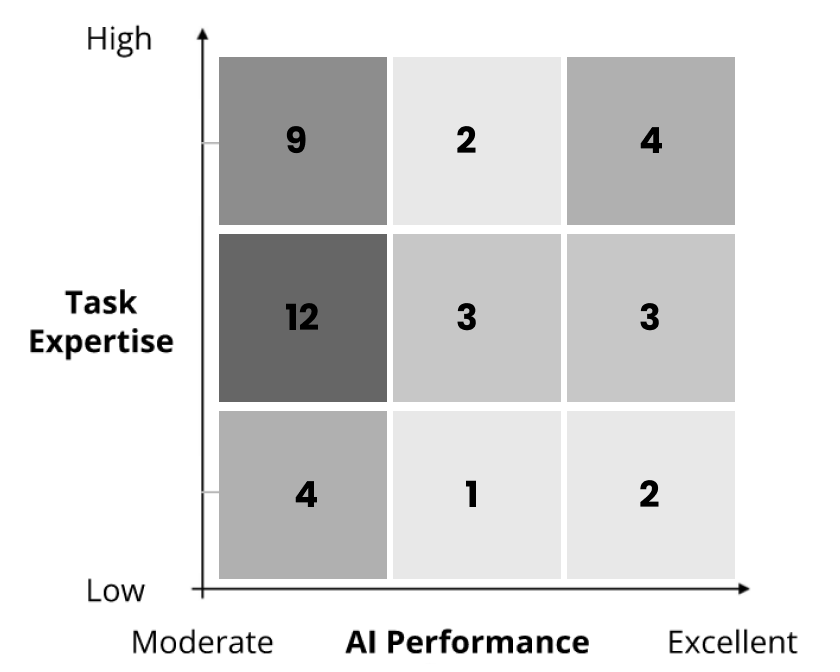}
    \caption{Yildirim et al.'s matrix}\cite{yildirim2023creating}
    \label{fig:mymatrix}
  \end{subfigure}
  \hfill
  \begin{subfigure}{0.47\textwidth}
    \centering
    \includegraphics[width=\linewidth]{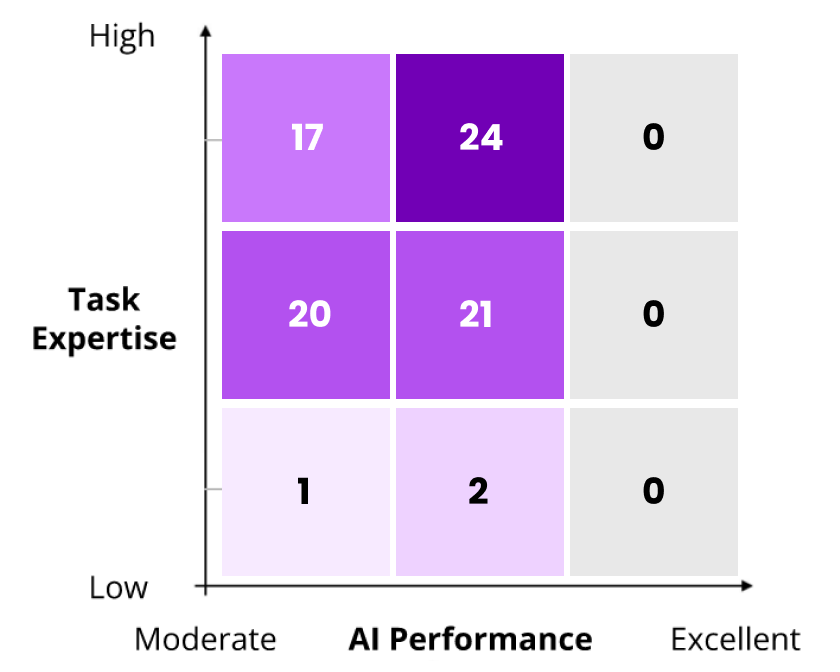}
 
    \caption{Our matrix: Applications with Pre-trained models}
    \label{fig:nurmatrix}
  \end{subfigure}
  \caption{Task-expertise and Model-performance}
  \label{fig:matrix}
\end{figure*}

We plotted the inferred task expertise and minimum model performance for each of the 85 applications (Figure \ref{fig:matrix}). We were surprised to see that none of the applications required excellent model performance to create value for their users. In addition, we were surprised that only three of the 85 applications had a task expertise less than a typical adult. These three included a system that logs everyday activities, a semantic image searching tool, and a chatbot designed to reduce loneliness \cite{cai2024pandalens, barnaby2024photoscout, xygkou2024mindtalker}. The majority of applications spread evenly from the upper-left corner (expert task/moderate performance) to the center (typical adult task/good performance).

We compared our matrix to the one resulting from an analysis of narrow AI \cite{yildirim2023creating}(Figure \ref{fig:mymatrix} and Figure \ref{fig:nurmatrix}). The narrow AI matrix had more density in the moderate model performance sector. It also had more density for less than a typical adult task expertise and more density for applications with excellent model performance. Examples of narrow AI applications with excellent model performance included things like medical imaging analysis (expert task/excellent performance) and biometric security (less than adult task/excellent performance). None of the applications we analyzed required excellent model performance. Narrow AI applications that required a level of expertise less than a typical adult included IoT sensing systems like smartwatch workout detection (less than adult expertise/moderate performance) and simple two-class classifiers like the biometric security. The applications we analyzed did not use pre-trained models for processing low-level sensor data, nor did they focus on simple two-class classification tasks.

\subsection{Interaction Design Patterns}
\begin{table*}[]
\caption{Emerging Interaction Design Patterns from the research applications.}
\label{tab:patterns}
\footnotesize 

\setlength{\tabcolsep}{4pt} 

\begin{tabular}{p{3cm} p{5cm} p{5cm} p{3cm}} 
\toprule
\textbf{Interaction Design Pattern} & \textbf{Problem} & \textbf{Solution}& \textbf{Artifact Number} \\
\midrule
\textbf{Chatbot Interview} & Need to collect customer information & Collects the needed information by having a chatbot interview & 2, 4, 5, 9, 11, 12, 21, 22, 23, 26, 34, 35, 40, 41, 42, 43, 45, 48, 49, 51, 53, 54, 55, 60, 61, 73 \\
\midrule
\textbf{Reveal Dimensions} & Uncertain about which dimensions are important for content creation and decision-making & Surfaces the dimensions others have used or previous users found useful. & 1, 2, 4, 5, 9, 10, 12, 13, 17, 18, 19, 24, 26, 29, 30, 32, 33, 34, 37, 43, 44, 45, 48, 50, 63, 66, 68, 69, 78, 79, 80 \\
\midrule
\textbf{Something Like This} & Struggle to communicate their desire in words & Express desires through examples & 3, 13, 16, 18, 38, 67, 80 \\
\midrule
\textbf{Dessert Cart} & Lack clarity about their overall desires. Uncertain about what they want & Provide multiple versions of what they are looking for & 1, 3, 4, 5, 9, 10, 13, 15, 16, 18, 23, 25, 26, 27, 30, 31, 32, 33, 35, 39, 47, 48, 51, 52, 57, 60, 62, 67, 68, 71, 77, 79, 80, 85 \\
\midrule
\textbf{Refine This} & Lack specificity regarding the details of their desires. Unsure of how to improve something. & Guides users in refining their ideas in real-time & 5, 6, 9, 13, 15, 16, 17, 18, 24, 26, 28, 29, 30, 35, 37, 51, 52, 57, 58, 62, 67, 75, 79, 81, 82, 83\\
\midrule
\textbf{Complete This} & Struggle to complete a communication task & Provides updated drafts and guidance to help users complete content creation tasks & 6, 10, 14, 15, 16, 17, 19, 24, 30, 34, 35, 52, 56, 59, 62, 63, 65, 66, 76\\
\midrule
\textbf{Blank Page Paralysis} & Feel overwhelmed by a blank page and are unsure how to start & Generates an initial draft, intermediate result, or suggested methods and materials & 2, 7, 8, 18, 22, 25, 28, 29, 33, 36, 38, 42, 43, 46, 50, 51, 58, 70, 72, 74, 75, 79 \\
\bottomrule
\end{tabular}
\end{table*}\label{tab:patterns}

Our analysis revealed seven emerging interaction design patterns used across these applications: ChatBot Interview, Reveal Dimensions, Something Like This, Dessert Cart, Refine This, Complete This, and Blank Page Paralysis (Table \ref{tab:patterns}). Many applications employed more than one of these patterns in their interaction designs. Below, we detail the seven patterns, discuss the interaction problem they address, and use examples to illustrate how the patterns work.

\subsubsection{\textbf{Chatbot Interview}}

Service providers often need detailed information from their customers in order to provide effective service. Today, they might use a complicated form or an interview at the start of a customer's journey to collect this information. The \textbf{\textit{ChatBot Interview}} pattern collects the needed information by having a ChatBot interview the user.
\par
\begin{figure*}
\centering

\begin{subfigure}{0.38\textwidth}
  \centering
  \includegraphics[width=0.7\linewidth]{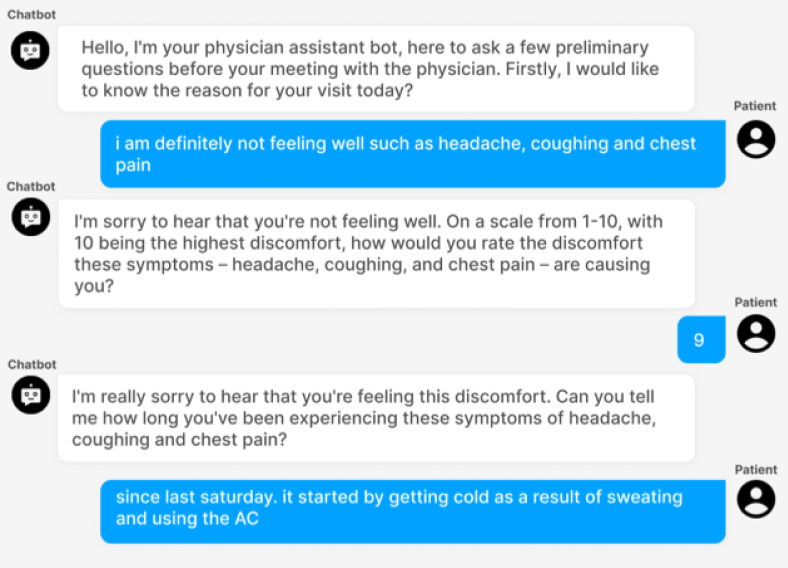}

  \caption{Pre-Consultation (Artifact Number: 64, \textit{CHI' 24})}
  \label{fig:preconsultant}
\end{subfigure}
\hspace{0.01\textwidth}
\begin{subfigure}{0.42\textwidth}
  \centering
  \includegraphics[width=0.7\linewidth]{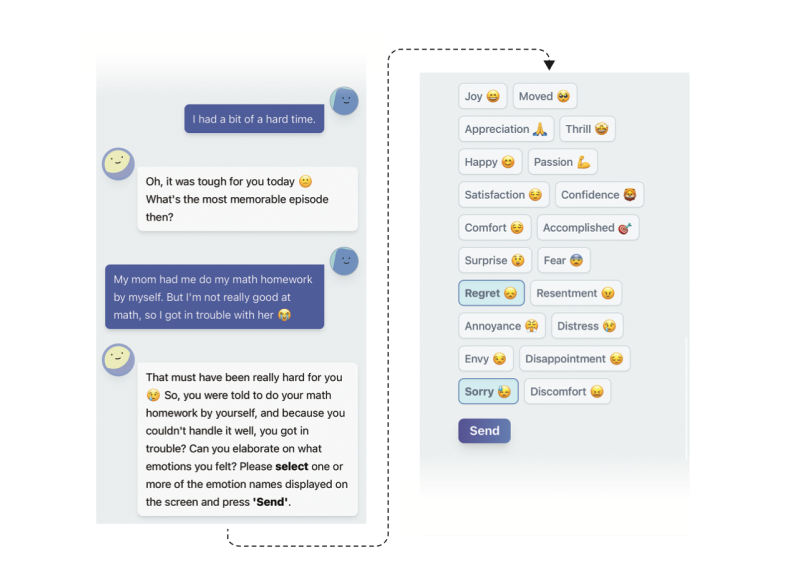}

  \caption{CHACHA Chatbot (Artifact Number: 23, \textit{CHI' 24})}
  \label{fig:chacha}
\end{subfigure}

\caption{Examples of Chatbot Interview}
\label{fig:chatinterview}
\end{figure*}

\textbf{\textit{Examples from Our Corpus }}
The Pre-Consultation application uses a ChatBot to interview patients in the waiting room (Figure \ref{fig:preconsultant})\cite{li2024beyond} . It collects medical history needed for an effective patient-clinician interaction. It replaces a clinician or the use of a tablet-based form.  The CHACHA application interviews children to collect the emotions connected to different personal events (Figure \ref{fig:chacha}) \cite{seo2024chacha}. It includes features that facilitate emotional expression, including a peer-like persona. It works to avoid single word responses from the child. CHACHA is meant to overcome the current problem of consistent data that comes from the many different clinicians who currently conduct interviews.

\subsubsection{\textbf{Reveal Dimensions}}
dUsers often do not know what dimensions matter when they create content or making a decision. The \textbf{\textit{Reveal Dimensions}} pattern surfaces the dimensions others have used when completing the same or similar task. This operationalizes the concept of collective intelligence and leads users to better outcomes. The \textbf{\textit{Reveal Dimensions}} pattern synthesizes data and guides users in structured exploration.
\par
\begin{figure}
\centering
\begin{subfigure}{0.38\textwidth}
\centering
\includegraphics[width=0.7\linewidth]{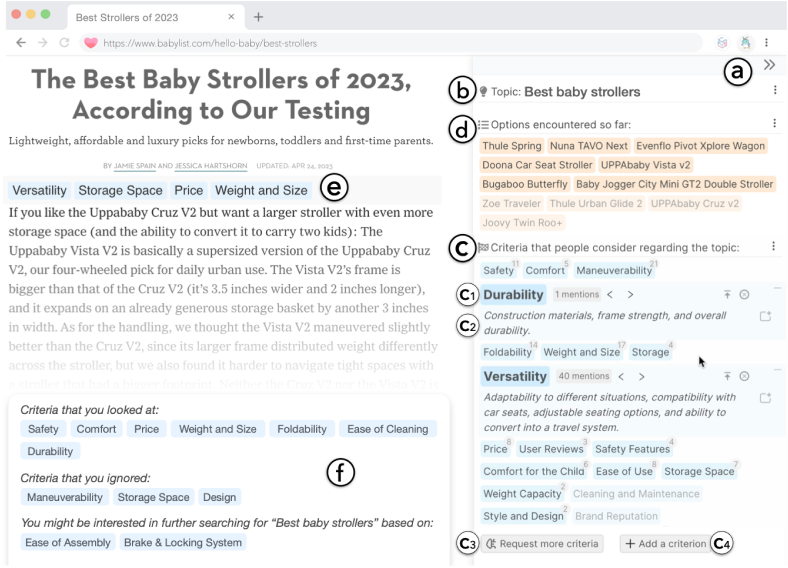} 
\caption{Selenite (Artifact Number: 1, \textit{CHI' 24})}
\label{fig:selenite}
\end{subfigure}
\hspace{0.01\textwidth}
\begin{subfigure}{0.38\textwidth }
\centering
\includegraphics[width=0.7\linewidth]{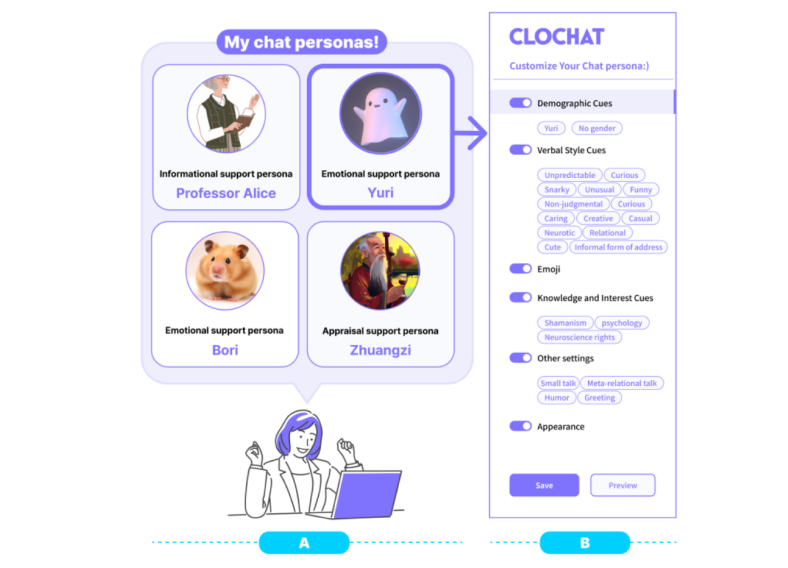}
\caption{CloChat (Artifact Number: 4, \textit{CHI' 24})}
\label{fig:clochat}
\end{subfigure}
\caption{Examples of Reveal Dimensions}
\label{fig:revealdimensions}
\end{figure}

 \textbf{\textit{Examples from Our Corpus :}}
The Selenite application helps users make a purchase decision for products that they are new to or unfamiliar with (Figure \ref{fig:selenite}) \cite{liu2024selenite}. It reveals the criteria commonly used by prior users when completing this task. The CloChat application (Figure \ref{fig:clochat}) \cite{ha2024clochat} supports users in designing the persona for a conversational agent. It provides users with six different factors, each with a set of options, helping them consider important features. By exploring these dimensions, users can better understand their preferences and more effectively create a conversational partner they want to use.

\subsubsection{\textbf{Something Like This}}

People sometimes want a new thing that is similar to something they are familiar with. For example, when going to get a haircut, a person might take an image of someone with the haircut they desire to communicate more effectively with a stylist. The \textbf{\textit{Something Like This}} pattern allows users to express what they want through examples, making it easier to convey their intentions. This pattern facilitates communication by enabling users to share visual or conceptual examples, such as a moodboard, that represent their desires.
\par
\begin{figure}[]
\centering
\begin{subfigure}{0.38\textwidth}
\centering
\includegraphics[width=0.7\linewidth]{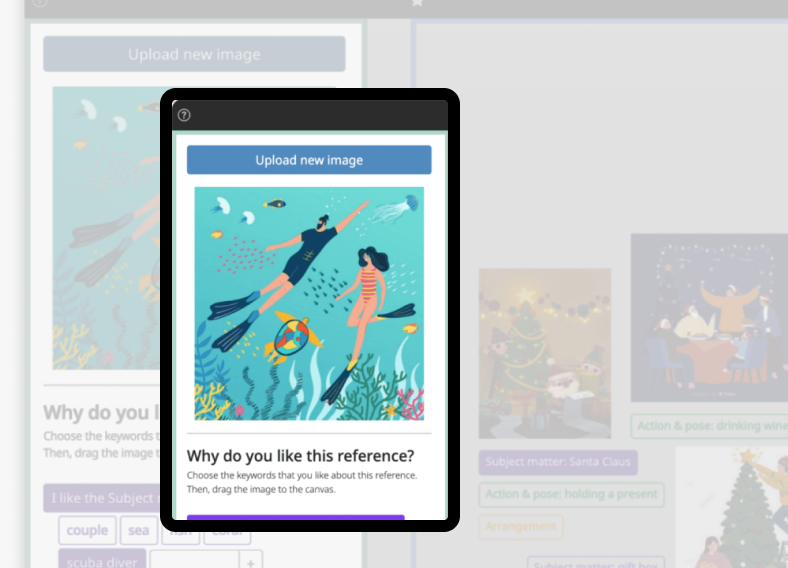} 
\label{fig:creativeconnect}
\end{subfigure}
\hspace{0.05\textwidth}
\begin{subfigure}{0.38\textwidth }
\centering
\includegraphics[width=0.7\linewidth]{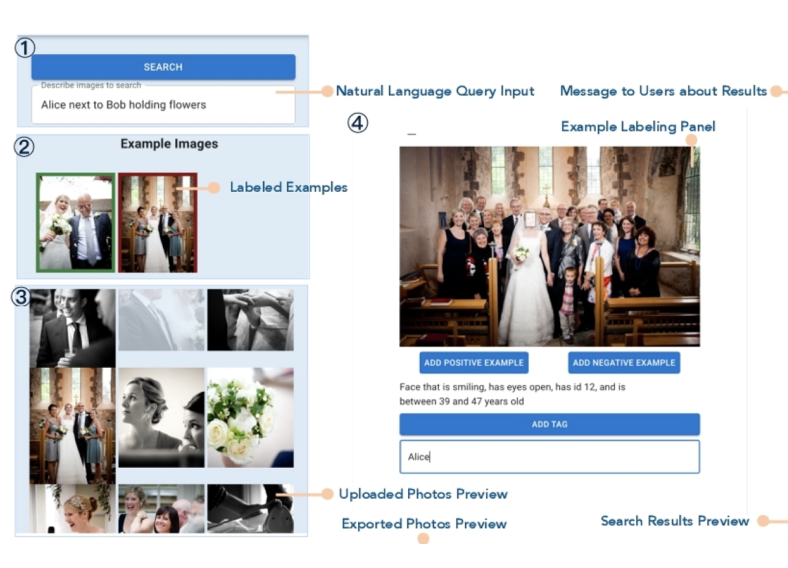}
\caption{Photoscout (Artifact Number: 38, \textit{CHI' 24})}
\label{fig:photoscout}
\end{subfigure}
\caption{Examples of Something Like This}
\label{fig:somethinglikethis}
\end{figure}

 \textbf{\textit{Examples from Our Corpus : }}
The CreativeConnect (Figure \ref{fig:creativeconnect}) \cite{choi2024creativeconnect} application supports designers' ideation by letting them share examples. The interface asks users to upload a reference image. It then  generates keywords and retrieves similar images that encourage users to refine their requests. PhotoScout (Figure\ref{fig:photoscout})\cite{barnaby2024photoscout} is a multi-modal image search tool which allow users can communicate their intention. To find the desired image, users provide a natural language prompt and refine the search by uploading both positive and negative example images.


\subsubsection{\textbf{Dessert Cart}}

When starting to create new content or to solve a problem, users sometimes only have a vague idea of what they want. They lack clarity on their desire. This makes it challenging to describe what they want. In many cases, users might know what they want when they see it. The \textbf{\textit{Dessert Cart}} pattern provides users with several versions (images, color palettes, stories, or documents) that they can choose from. By offering a variety of options, the \textbf{\textit{Dessert Cart}} pattern helps users hone their desire. 

\par
\begin{figure}[h]
\centering
\begin{subfigure}{0.38\textwidth}
\centering
\includegraphics[width=0.7\linewidth]{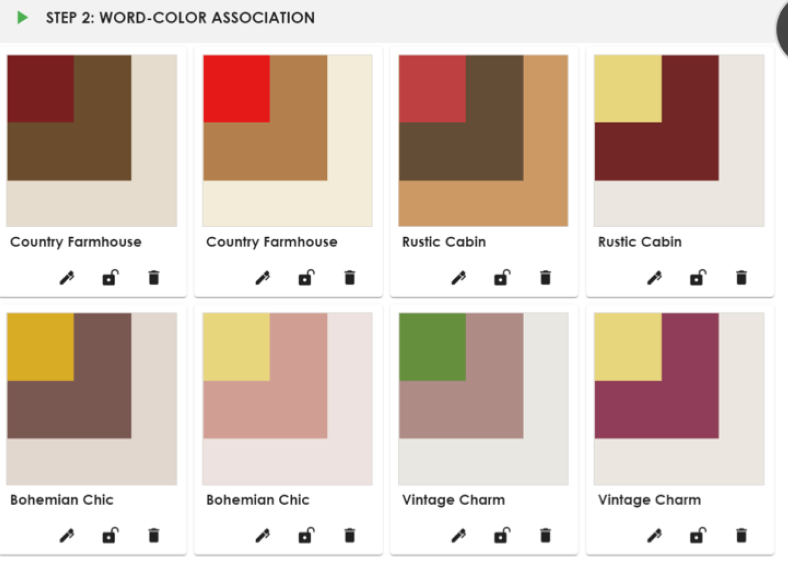}
\caption{C2Idea (Artifact Number: 26, \textit{CHI' 24})}
\label{fig:c2idea}
\end{subfigure}
\hspace{0.05\textwidth}
\begin{subfigure}{0.38\textwidth }
\centering
\includegraphics[width=0.7\linewidth]{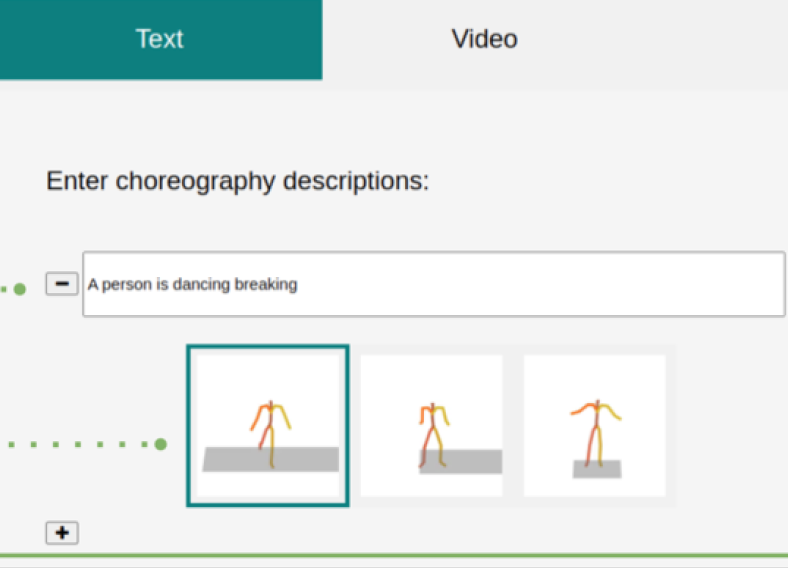}
\caption{DanceGen (Artifact Number: 68, \textit{DIS' 24})}
\label{fig:dancegen}
\end{subfigure}
\caption{Examples of Dessert Cart}
\label{fig:dessertcart}
\end{figure}

 \textbf{\textit{Examples from Our Corpus : }}
The C2Ideas application (Figure \ref{fig:c2idea}) \cite{hou2024c2ideas} helps users mock up a new interior design. Users provide a few keywords to express their desired mood and tone. C2Ideas generates eight color palettes based on the users' input. Users can select one or they can choose new keywords. The DanceGen application (Figure \ref{fig:dancegen}) \cite{liu2024dancegen} helps users create unique dance motions. Users describe what they want with words and DanceGen provides three options. By exploring the variations, users can move closer to their desire.

\subsubsection{\textbf{Refine This}}
Users often start with an idea or a general sense of what they want but lack clarity on the specific details. They may only have a high-level or aggregate understanding of their concept and feel uncertain about how their idea will look or function. The \textbf{\textit{Refine This}} pattern guides users in refining their ideas in real-time. This approach allows users to progressively narrow down their options, helping them identify their specific desires. By visualizing the changes and providing intermediate results, users can move closer to a well-defined outcome and clearly identify their desire.

\par
\begin{figure}[h]
\centering
\begin{subfigure}{0.36\textwidth}
\centering
\includegraphics[width=0.7\linewidth]{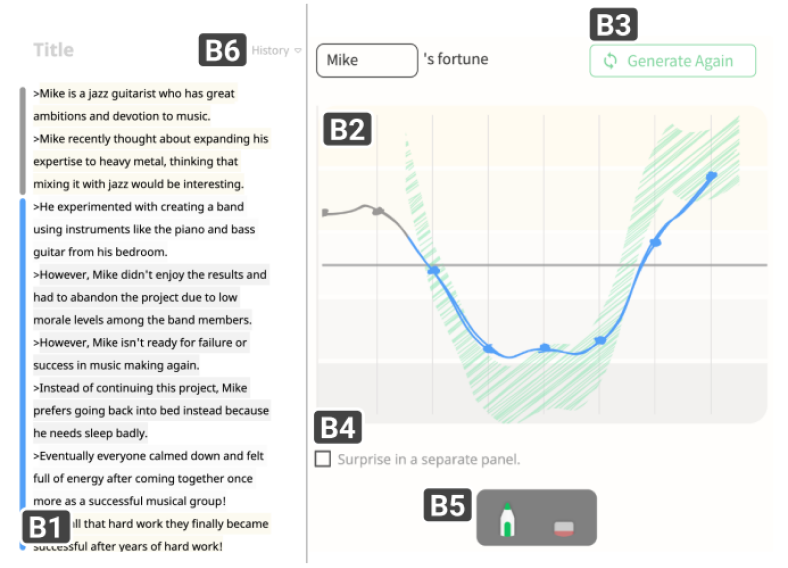} 
\caption{\centering Talebrush (Artifact Number: 81, \textit{CHI' 22})}
\label{fig:talebrush}
\end{subfigure}
\hspace{0.05\textwidth}
\begin{subfigure}{0.36\textwidth }
\centering
\includegraphics[width=0.7\linewidth]{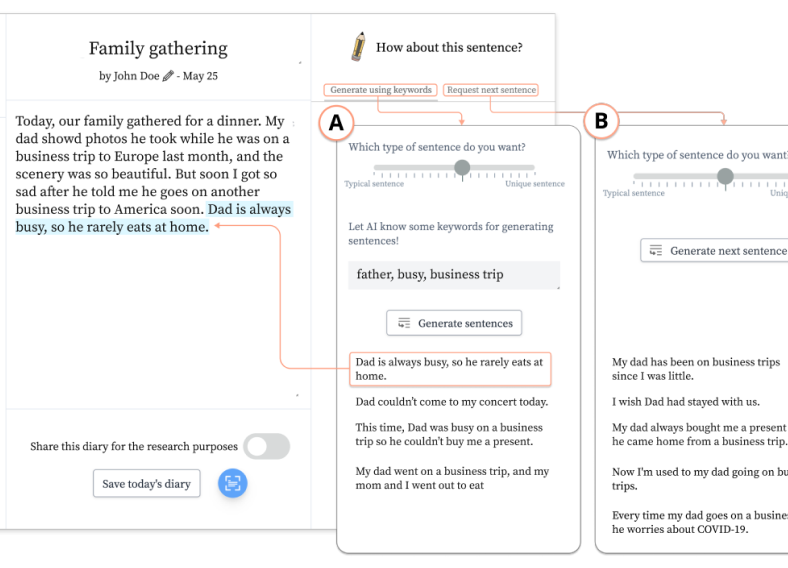}
\caption{\centering DiaryMate (Artifact Number: 52, \textit{CHI' 24})}
\label{fig:diarymate}
\end{subfigure}  
\caption{Examples of Refine This}
\label{fig:Refinethis}
\end{figure}

\textbf{\textit{Examples from Our Corpus : }}
TaleBrush (Figure \ref{fig:talebrush}) \cite{chung2022talebrush} is a story-generating tool that works by sketching a protagonist's fortune. While co-generating stories with TaleBrush, users draw a line plot of the character's fortune on the right side and can view the generated story in real time on the left side. After reading the story, users can refine it by re-sketching the fortune plot to make it more extreme or calm. DiaryMate (Figure \ref{fig:diarymate}) \cite{kim2024diarymate} is a personal journal writing assistant that suggests the next sentence for a journal. While users receive suggestions, they can view the LLM-generated sentences and adjust the type of sentence in real time. Users can change the tone of a sentence with sliders, and confirm the change in real time. 

\subsubsection{\textbf{Complete This}}

Users sometimes struggle to finish a content generation task because aspects of it feel overwhelming or laborious or they may feel uncertain about how to proceed. The \textbf{\textit{Complete This}} pattern addresses this challenge by providing updated drafts of the communication to guide users toward completion. 
\par
\begin{figure}[h]
\centering
\begin{subfigure}{0.38\textwidth}
\centering
\includegraphics[width=0.7\linewidth]{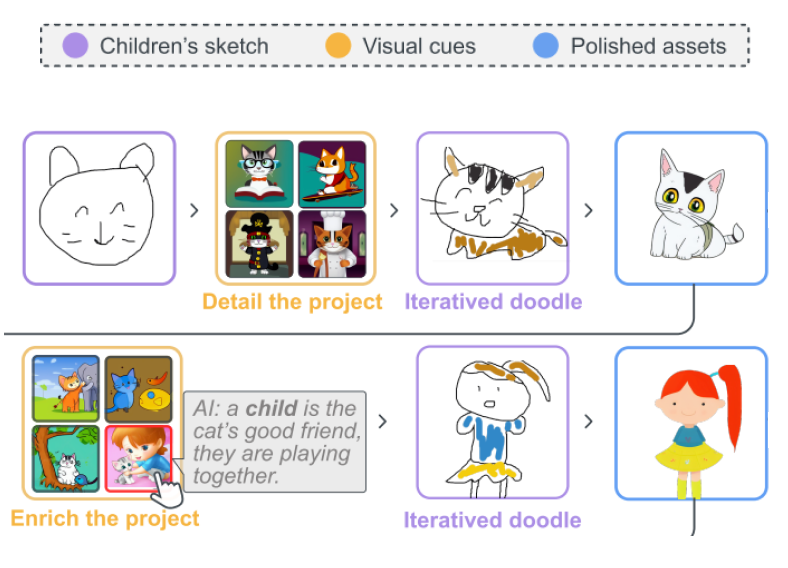} 
\caption{ChatScratch (Artifact Number: 35, \textit{CHI' 24})}
\label{fig:chatscratch}
\end{subfigure}
\hspace{0.05\textwidth}
\begin{subfigure}{0.38\textwidth }
\centering
\includegraphics[width=0.7\linewidth]{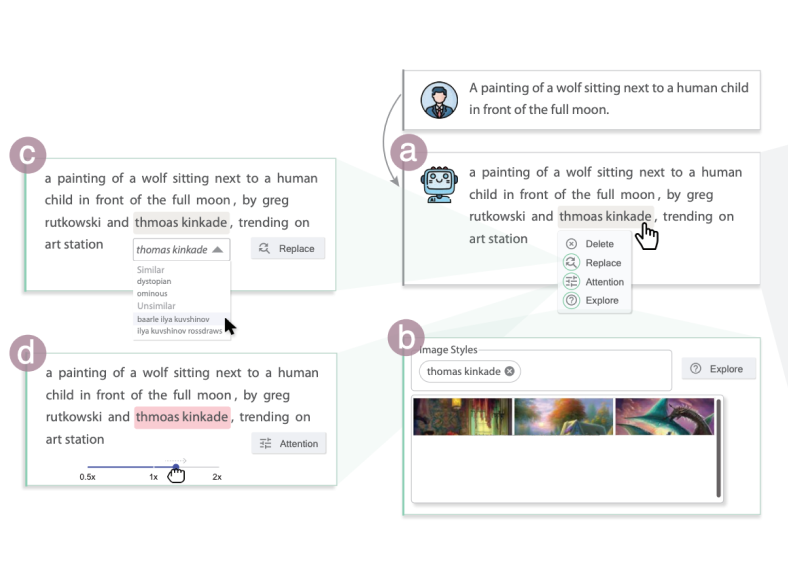}
\caption{PromptCharm (Artifact Number:15,\textit{CHI' 24})}
\label{fig:promptcharm}
\end{subfigure}
\caption{Examples of Complete This}
\label{fig:Completethis}
\end{figure}

\textbf{\textit{Examples from Our Corpus : }}
ChatScratch (Figure \ref{fig:chatscratch}) \cite{chen2024chatscratch} is a learning tool which teaches programming through interactive storyboards and digital drawings. Based on an initial sketch created by a child, ChatScratch creates a polished version. PromptCharm (Figure \ref{fig:promptcharm}) \cite{wang2024promptcharm} facilitates text-to-image creation through prompt engineering. Users write an accurate prompt that the computer can understand. Since writing proper prompts can be challenging, PromptCharm assists users by allowing them to refine their draft prompts. It then generates an image from the text prompt. 

\subsubsection{\textbf{Blank Page Paralysis}}

When creating content, users often feel overwhelmed by the blank page and are unsure of how to start. The \textbf{\textit{Blank Page Paralysis}} pattern addresses this challenge by generating an initial draft, intermediate result, or suggested methods and materials. It provides users with a starting point to respond to, helping them overcome the lack of inertia from the blank page.
\par
\begin{figure}[]
\centering
\begin{subfigure}{0.36\textwidth}
\centering
    \includegraphics[width=0.7\linewidth]{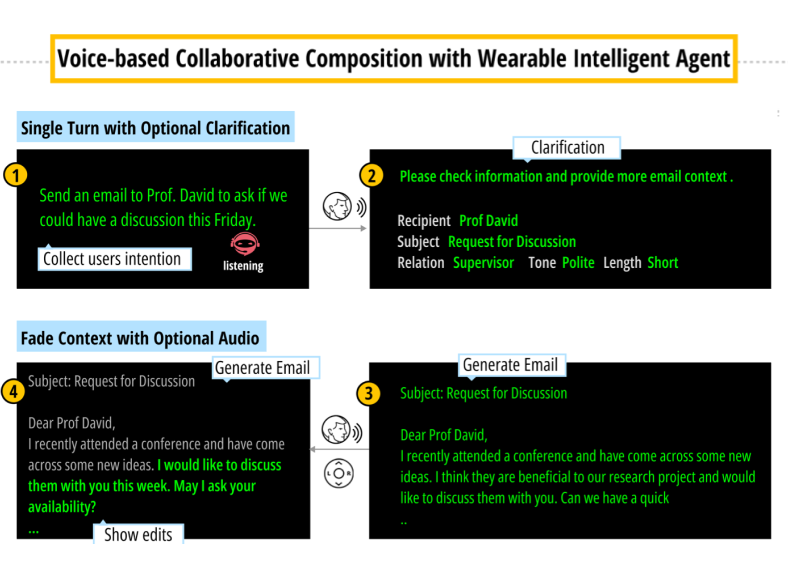}
    \caption{\centering GlassMail (Artifact Number: 70, \textit{DIS' 24})}
    \label{fig:glassmail}
\end{subfigure}
\hspace{0.05\textwidth}
\begin{subfigure}{0.36\textwidth}
\centering
    \includegraphics[width=0.7\linewidth]{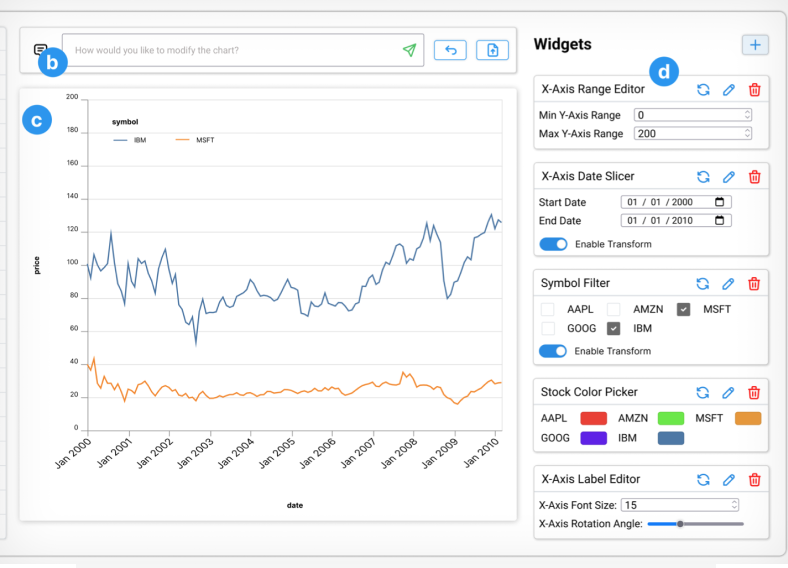}
    \caption{DynaVis (Artifact Number: 29, \textit{CHI' 24})}
    \label{fig:dynavis}
\end{subfigure}
\caption{Examples of Blank Page Paralysis}
\label{fig:Blankpage}
\end{figure}

\textbf{\textit{Examples from Our Corpus : }}
The GlassMail application (Figure \ref{fig:glassmail}) \cite{zhou2024glassmail} is an email creation assistant. GlassMail generates a draft of an email and then asks the user to review and edit. The DynaVis application (Figure \ref{fig:dynavis}) \cite{vaithilingam2024dynavis} provides both natural language input and UI widgets to help users create a visualization. Users can provide natural language commands to edit a visualization. The system first generates a default visualization along with controls based on the user's imported data. 

\section{Discussion}
Innovators often want to know where products and services are currently co-creating value for customers and service providers to help them select a lower risk starting place for innovating. This is particularly challenging for AI innovators who want to leverage the capabilities of pre-trained models, where commercial examples are caught in a hype cycle from venture capital funding that makes inferring where success is happening quite difficult. To help them, we developed a resource of \textquotedblleft successful\textquotedblright applications by analyzing recent HCI research applications. We analyzed 85 applications, identifying the domains, model performance, and task expertise for each. We extracted 294 pre-trained model capabilities, and we documented seven emerging design patterns. We view these resources-including the set of applications with their domains, model performance, and task expertise; the collection of pre-trained model capabilities; and the small set of emerging design patterns- as \textquotedblleft \textit{better than nothing},\textquotedblright the current state of the world for innovators who want to create new things with pre-trained models. Below we discuss the implications and limitations of our research.

\subsection{Value in Understanding Content}
Many people refer to pre-trained models as Generative AI or GenAI. Thus, it is easy to assume innovators should focus on using pre-trained models to automate content generation.  Most of the applications we analyzed create value for users by helping them generate or improve new artifacts made up largely of text and images. What surprised us was the much larger number of capabilities that deliver content understanding. 

Researchers leveraged content understanding as a step towards generating summaries, answering questions, finding similar things, creating rankings/recommendations, classifying and extracting specific elements and items, refining and improving content, and interpreting content's meaning in order to draw out subtextual insights. In hindsight, it seems obvious that in order to generate content, a system would also need some level of content understanding. We may have overlooked this due to NLP research that has historically separated generation (NLG) and understanding (NLU). We were impressed with researchers' ability to effectively bring these aspects together. We view content understanding as a potential starting place for innovation and as a topic for additional HCI research. An open challenge for both researchers and innovators is addressing the moderate to good level of model performance. How should researchers or innovators discover opportunities where moderate to good levels of content understanding might be valuable and useful? This seems like a ripe target for new research on AI innovation methods.

\subsection{Revealing Gaps}
In reflecting on the applications, we began to notice important gaps. We identified two areas that seem underinvestigated. They both have great potential for co-creating user and service provider value. First, applications rarely ever focus on producing many similar artifacts for large audiences. Second, applications only touched a small number of domains.

Manufacturing creates value by making it fast and cheap to create the identical things for many people. For example, a fast fashion brand can design a pair of jeans and manufacture thousands of pairs to sell globally. At the other end of a continuum, crafts create value by having a craftsperson make a one-of-a-kind product that is carefully handcrafted. This would be like a high-fashion brand design tailored jeans for a celebrity. Every jeans will be unique and different, but each takes a lot of effort. Pre-trained models have the potential to disrupt the space between the two ends of this continuum. Pre-trained models hold the promise to craft unique artifacts for each person. Using the fashion analogy, a designer could create a standard jean, and pre-trained models could make individual versions for each customer. This might include changing body types or climates, or some other quality the customer cares about. This is something manufacturing cannot do and that is expensive for craft to do.

The applications we analyzed almost exclusively focused on helping a single user make a single thing. They help a homeowner generate an interior design, help a researcher find a good research question, or help a designer create the perfect ChatBot persona. However, none of the applications investigated how pre-trained models might be used to create many similar versions of the same thing tailored to different individuals or groups. This overlooks the broader applicability of a single application, which could be valuable to numerous homeowners facing diverse design challenges. In reflecting on this, we quickly thought of 100 plus examples where value might be co-created by making many things for many people. A job searcher might use their standard cover letter and resume along with a set of job listings to customize their materials for each job application. An advertising company might write an email marketing copy for a new product and then use descriptions of different customer segments to create many targeted versions of the ad. A font designer might create a few letter forms and then get pre-trained models to generate the other letters, different weights, italics, and ligatures. Pre-trained models should be able to automate much more personalized production, but researchers have not explored this. It could be that researchers did develop these sorts of applications, but they were not accepted for publication, thus they were not a part of our analysis.

The applications  in our corpus only touched on a small number of domains. Researchers did not make applications for manufacturing, agriculture, transportation, government, or finance, even though narrow AI applications have created value for these domains. Industry media shows that the banking and the financial services sector have traditionally been one of the fastest adopters of new technology. Journalists note that this domain is playing with new services that utilize pre-trained models, such as intelligent customer support \cite{FutureProofing}. Interestingly, none of the applications in our corpus touched on this domain. The small number of domains explored might say more about the inchoate state of HCI research on pre-trained models or on the partnerships and collaborations HCI researchers currently have. Given this unexpected gap, we encourage researchers to explore these less investigated domains. 

It was less surprising to see that across the capabilities described by Yildirim et al. \cite{yildirim2023creating}, that optimization and forecasting, which seem to play a large role in narrow AI, did not appear in the applications researchers developed. We saw nothing close to well known and very successful AI capabilities such as predictive maintenance, demand prediction (smart warehousing), or digital twins, which help companies prototype new ways to optimize. We suspect that we did not see these things because they are not central to the more generative capabilities that seem to dominate pre-trained models. We do not claim optimization and forecasting are not possible with pre-trained models, only that HCI researchers do not seem to be trying to get them to take on these traditional AI strengths.

\subsection{How Innovators Might Use Capabilities and Interaction Design Patterns}
\begin{figure*}[b]
  \centering
  \includegraphics[width=0.9\linewidth]{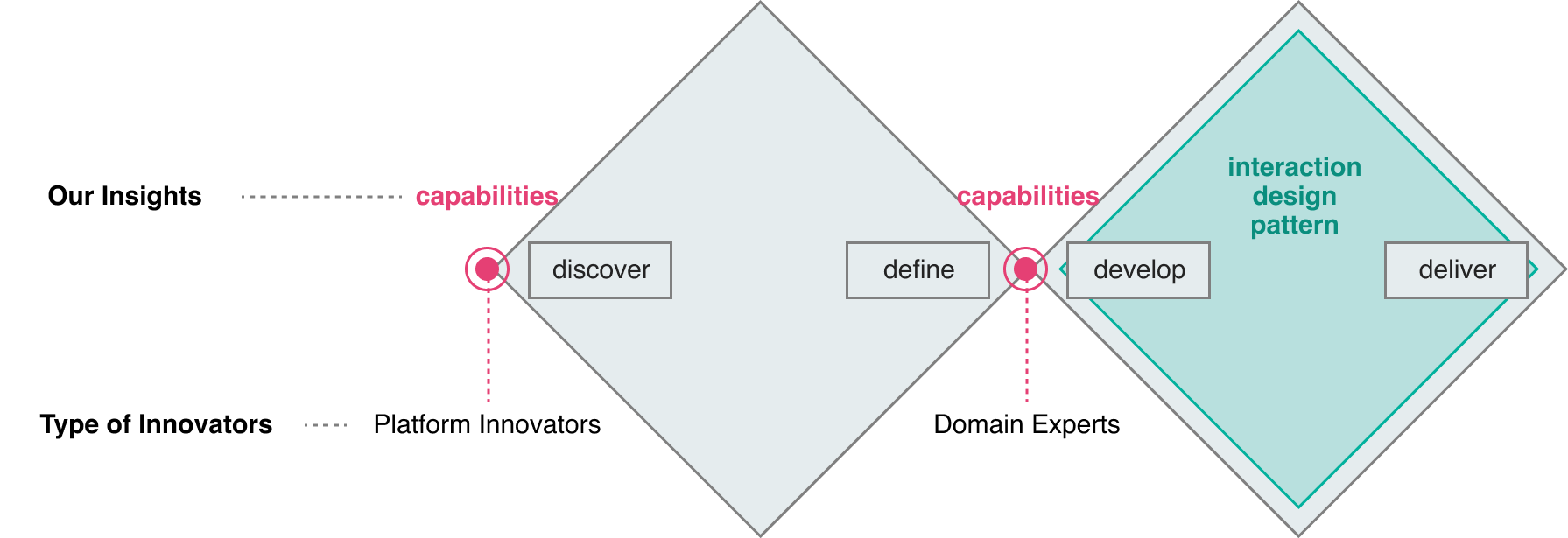}
  \caption{ The overview of when our insights can be useful to innovators. Yang et al,\cite{yang2020re} claimed AI innovation starts at the center of double diamond.}
  \label{fig:discussion}
\end{figure*}

We believe the resources we developed using research applications as a proxy for commercial success can be useful to innovators. Below we share a few ways they might engage with these resources that emerged from our reflection on the research. We make use of the familiar double diamond, user-centered innovation model to help illustrate opportunities for use. We discuss usage in terms of \textit{domain experts} (innovators with expertise in an industrial sector such as healthcare, finance, hospitality) and \textit{platform innovators} (innovators with expertise in a technology platform such as mobile, social media, web application, robotics, VR)  (Figure \ref{fig:discussion}).

Innovators working as domain experts might benefit from thinking about pre-trained model capabilities when they are at the center of the double diamond. This assumes they have a good understanding of their customers' and users' needs. They draw on this knowledge as they search for new things to make that co-create value. Their goal is to find a harmonious intersection of needs and reasonable AI capabilities. They can look over our set of capabilities and ask themselves; When might my customers or users might find this useful? What activities are they engaged in where this might be useful? Next, they can explore if fulfilling a need with one of the capabilities would be valuable. Would it create more value than costs. Note, Yang et al., \cite{yang2020re} described domain driven AI innovation as starting in the center of the double diamond. They start with the selection of an application or feature they think will help users.

Platform innovators might make use of the capabilities and domains at the start of the double diamond. They have deep knowledge of the  platform they typically use when innovating. They could review the list of capabilities to envision matches, where the capability extends what their platform can already do. These innovators can  use matchmaking \cite{bly1999design}, a technology-centered approach to innovation, which starts with capabilities and then searches for the best customer-to-application pair. For example, if someone invented velcro, they would use matchmaking to search for your best customer. Who needs velcro? Who has such a strong need they might pay the most for this capability? HCI researchers have previously had luck using matchmaking or hybrid matchingmaking and user-centered design process to innovate with AI \cite{liu2024human, yildirim2023creating}.

The design patterns might be useful to any type of innovator when they reach the middle of the second diamond and start to iteratively improve a prototype, turning into a real product or service. Innovators can look at the different interaction design challenges they face and review the patterns to see if one or more might help resolve this challenge.

\section{Limitations}
Our work has two limitations. First, our research and of the resources we developed for innovators comes from our choice to exclusively focus on pre-trained model applications from researchers who publish at ACM CHI and DIS. We could have analyzed commercial applications; however, the current hype cycle surround pre-trained models made us view this source as less valuable than research applications. We could have included research using pre-trained models more broadly, from a larger set of HCI and AI researchers. However, we chose to focus on CHI and DIS as it brings with it a concern for user needs and an emerging understanding of the issues and risks around responsible AI. We feel our choice to only analyze research applications makes our resources incomplete, but not incorrect. Second, our research heavily relies on a single prior work—specifically, Yildirim et al.'s study on AI capabilities, particularly in identifying domains and structuring the capability analysis. While this foundational work is highly relevant, such reliance may have shaped or constrained our findings. Therefore, we view these resources as preliminary and hope that future researchers will contribute new capabilities, additional insights on model performance and expertise, and many more interaction design patterns.

Innovators want to know where they are likely to experience success. Our use of the term \textquotedblleft commercially successful\textquotedblright is meant to capture products and services that have a long history of value creation and success. Things like \textit{restaurants} have been around for a long time.  There is  a market, a collection of people who regularly go out and eat. Creating a new restaurant does not require users to adopt some new behavior. The success of restaurants does not mean that every new restaurant will experience success, but it implies that trying to create a new restaurant is less risky than trying to create things that require larger changes in people's behaviors. We do not see any of the current, publicly available applications that leverage pre-trained models as commercially successful because they do not have evidence from many, many years of success. Most of the applications are funded by investors who hope for future commercial success. We recognize that the HCI research corpus has a huge blind spot in terms of financial risks. Financial success is almost never an HCI research focus. However, we worried that a corpus based on things investors have chosen to fund would bring a host of hidden issues with respect to responsible AI and to real user needs.

\section{Conclusion}
Our work provides an overview of what is working with pre-trained models. From previous cases of innovation, we see that understanding what works with specific technology allows innovators to mitigate risk by providing a safe opportunities for innovation. To this end, we examined applications built with pre-trained models, analyzing their domains and data types, exploring their capabilities, and assessing their model performance and task expertise. Our exploration reveals that pre-trained models have significant potential and value in understanding content, and we have identified unexplored opportunity spaces for their use. Throughout our investigation, we identified seven high-level interaction design patterns that could play a crucial role in bridging the gap for innovators. We advocate that our exploration of what is working with pre-trained models can provide valuable insights into finding an appropriate starting point for designing with these models, and envision new ways to innovate with cutting-edge pre-trained models.

\begin{acks}
We would like to thank Nik Martelaro, Ken Holstein, Kyzyl Monteiro and Faria Huq for their manuscript feedback. Finally, we would like to thank our anonymous reviewers for their feedback.
This research was partially funded by the National Science Foundation under Grant 2007501 and by a grant from Bloomberg.
\end{acks}


\bibliographystyle{ACM-Reference-Format}
\bibliography{main}


\appendix
\onecolumn
\section{The full list of 85 artifacts}

\begin{longtable}{p{0.1\textwidth} p{0.1\textwidth} p{0.4\textwidth} p{0.1\textwidth} p{0.1\textwidth}}
\caption{The full list of 85 artifacts}
\label{tab:artifactlist} \\

\toprule
\textbf{Artifact} & \textbf{Venue} & \textbf{Title} & \textbf{Model} & \textbf{Task} \\
\textbf{Number} &  &  & \textbf{Performance} & \textbf{Expertise} \\
\midrule
\endfirsthead

\toprule
\textbf{Artifact} & \textbf{Venue} & \textbf{Title} & \textbf{Model} & \textbf{Task} \\
\textbf{Number} &  &  & \textbf{Performance} & \textbf{Expertise} \\
\midrule
\endhead

\midrule
\multicolumn{5}{r}{{Continued on next page}} \\
\bottomrule
\endfoot

\bottomrule
\endlastfoot
1          & CHI'24           & Selenite: Scaffolding Online Sensemaking with Comprehensive Overviews Elicited from Large Language Models       & 1 & 2      \\

2           &CHI'24           & ABScribe: Rapid Exploration of Multiple Writing Variations in Human-AI Co-Writing Tasks using Large Language Models   & 1 & 2          \\

3          &CHI'24            & PaperWeaver: Enriching Topical Paper Alerts by Contextualizing Recommended Papers with User-collected Papers     & 1 & 2        \\

4          & CHI'24           & CloChat: Understanding How People Customize, Interact, and Experience Personas in Large Language Models      & 1 & 2       \\

5          &CHI'24         & Luminate: Structured Generation and Exploration of Design Space with Large Language Models for Human-AI Co-Creation    & 1 & 2        \\

6          &CHI'24           & The HaLLMark Effect: Supporting Provenance and Transparent Use of Large Language Models in Writing with Interactive Visualization   & 1 & 2       \\

7          &CHI'24          & Memoro: Using Large Language Models to Realize a Concise Interface for Real-Time Memory Augmentation    & 2 & 3         \\

8           &CHI'24          & Think Fast, Think Slow, Think Critical: Designing an Automated Propaganda Detection Tool   & 2 & 3          \\

9           &CHI'24          & Beyond Numbers: Creating Analogies to Enhance Data Comprehension and Communication with Generative AI  & 1 & 2        \\

10           &CHI'24          & CodeAid: Evaluating a Classroom Deployment of an LLM-based Programming Assistant that Balances Student and Educator Needs   & 2 & 3       \\

11          &CHI'24          & Care-Based Eco-Feedback Augmented with Generative AI: Fostering Pro-Environmental Behavior through Emotional Attachment    & 2 & 3         \\

12           &CHI'24          & CharacterMeet: Supporting Creative Writers' Entire Story Character Construction Processes Through Conversation with LLM-Powered Chatbot Avatars   & 2 & 2        \\

13          &CHI'24          & CreativeConnect: Supporting Reference Recombination for Graphic Design Ideation with Generative AI    & 2 & 3       \\

14          &CHI'24          & From Paper to Card: Transforming Design Implications with Generative AI    & 2 & 2       \\

15         & CHI'24          & PromptCharm: Text-to-Image Generation through Multi-modal Prompting and Refinement    & 1 & 3       \\

16          &  CHI'24          & RoomDreaming: Generative-AI Approach to Facilitating Iterative, Preliminary Interior Design Exploration    & 1 & 3       \\

17        &  CHI'24          & Jigsaw: Supporting Designers to Prototype Multimodal Applications by Chaining AI Foundation Models    & 2 & 3        \\

18          & CHI'24          & TypeDance: Creating Semantic Typographic Logos from Image through Personalized Generation     & 1 & 3        \\

19         & CHI'24          & Metamorpheus: Interactive, Affective, and Creative Dream Narration Through Metaphorical Visual Storytelling      & 1 & 3      \\

20          & CHI'24          & Mathemyths: Leveraging Large Language Models to Teach Mathematical Language through Child-AI Co-Creative Storytelling     & 2 & 2       \\

21          &  CHI'24          & Understanding the Impact of Long-Term Memory on Self-Disclosure with Large Language Model-Driven Chatbots for Public Health Intervention   & 2 & 2        \\

22        &  CHI'24          & MindfulDiary: Harnessing Large Language Model to Support Psychiatric Patients' Journaling     & 1 & 2      \\

23         &  CHI'24          & ChaCha: Leveraging Large Language Models to Prompt Children to Share Their Emotions about Personal Events     & 2 & 2       \\

24          &  CHI'24          & EvalLM: Interactive Evaluation of Large Language Model Prompts on User-Defined Criteria        & 2 & 3      \\

25        &  CHI'24          & VIVID: Human-AI Collaborative Authoring of Vicarious Dialogues from Lecture Videos      & 1 & 2      \\

26          &  CHI'24          & C2Ideas: Supporting Creative Interior Color Design Ideation with a Large Language Model   & 1 & 3  \\

27          &  CHI'24          & ChainForge: A Visual Toolkit for Prompt Engineering and LLM Hypothesis Testing       & 1 & 2       \\

28          &  CHI'24          & CoPrompt: Supporting Prompt Sharing and Referring in Collaborative Natural Language Programming   & 1 & 2          \\

29          &  CHI'24          & DynaVis: Dynamically Synthesized UI Widgets for Visualization Editing      & 2 & 3        \\

30         &  CHI'24          & PlantoGraphy: Incorporating Iterative Design Process into Generative Artificial Intelligence for Landscape Rendering       & 1 & 3       \\

31        &   CHI'24          & CollabCoder: A Lower-barrier, Rigorous Workflow for Inductive Collaborative Qualitative Analysis with Large Language Models   & 1 & 2          \\

32         &  CHI'24          & Farsight: Fostering Responsible AI Awareness During AI Application Prototyping    & 2 & 2          \\

33         &  CHI'24          & Generating Automatic Feedback on UI Mockups with Large Language Models     & 2 & 3         \\

34         &  CHI'24          & Teach AI How to Code: Using Large Language Models as Teachable Agents for Programming Education    & 1 & 2           \\

35          &  CHI'24          & ChatScratch: An AI-Augmented System Toward Autonomous Visual Programming Learning for Children Aged 6-12      & 2 & 3      \\

36         &  CHI'24          & AXNav: Replaying Accessibility Tests from Natural Language      & 2 & 3        \\

37          &   CHI'24          & Learning Agent-based Modeling with LLM Companions: Experiences of Novices and Experts Using ChatGPT and NetLogo Chat      & 1 & 3      \\

38         &  CHI'24          & PhotoScout: Synthesis-Powered Multi-Modal Image Search      & 1 & 1      \\

39          &  CHI'24          & RELIC: Investigating Large Language Model Responses using Self-Consistency     & 1 & 3                 \\

40        &   CHI'24          & Rehearsal: Simulating Conflict to Teach Conflict Resolution       & 1 & 2               \\

41          &  CHI'24          & VAL: Interactive Task Learning with GPT Dialog Parsing      & 2 & 2          \\

42         &   CHI'24          & VirtuWander: Enhancing Multi-modal Interaction for Virtual Tour Guidance through Large Language Models    & 2 & 2     \\

43        &   CHI'24          & A Piece of Theatre: Investigating How Teachers Design LLM Chatbots to Assist Adolescent Cyberbullying Education           & 1 & 3           \\

44          &  CHI'24          & MindShift: Leveraging Large Language Models for Mental-States-Based Problematic Smartphone Use Intervention            & 1 & 2          \\

45          &   CHI'24          & MindTalker: Navigating the Complexities of AI-Enhanced Social Engagement for People with Early-Stage Dementia        & 2 & 1       \\

46        &   CHI'24          & PANDALens: Towards AI-Assisted In-Context Writing on OHMD During Travels      & 2 & 1  \\

47        &  CHI'24          & Intelligent Support Engages Writers Through Relevant Cognitive Processes      & 1 & 2         \\

48          &  CHI'24          & BIDTrainer: An LLMs-driven Education Tool for Enhancing the Understanding and Reasoning in Bio-inspired Design       & 2 & 3               \\

49         &  CHI'24          & ClassMeta: Designing Interactive Virtual Classmate to Promote VR Classroom Participation         & 2 & 2      \\

50          &  CHI'24          & Co-Designing QuickPic: Automated Topic-Specific Communication Boards from Photographs for AAC-Based Language Instruction     & 2 & 2                 \\

51         &  CHI'24          & ContextCam: Bridging Context Awareness with Creative Human-AI Image Co-Creation       & 2 & 3      \\

52         &  CHI'24          & DiaryMate: Understanding User Perceptions and Experience in Human-AI Collaboration for Personal Journaling       & 1 & 2        \\

53         &  CHI'24          & See Widely, Think Wisely: Toward Designing a Generative Multi-agent System to Burst Filter Bubbles        & 2 & 2      \\

54         &  CHI'24          & SimUser: Generating Usability Feedback by Simulating Various Users Interacting with Mobile Applications      & 2 & 3     \\

55        &  CHI'24          & Open Sesame? Open Salami! Personalizing Vocabulary Assessment-Intervention for Children via Pervasive Profiling and Bespoke Storybook Generation         & 2 & 2      \\

56        &  CHI'24          & Unblind Text Inputs: Predicting Hint-text of Text Input in Mobile Apps via LLM      & 2 & 2      \\

57        &   CHI'24          & Prompting for Discovery: Flexible Sense-Making for AI Art-Making with Dreamsheets      & 1 & 3         \\

58          &   CHI'24          &  Rambler: Supporting Writing With Speech via LLM-Assisted Gis tManipulation      & 2 & 2         \\

59         &   CHI'24          & Marco: Supporting Business Document Workflows via Collection-Centric Information Foraging with Large Language Models      & 2 & 2        \\

60          &   CHI'24          & Scientific and Fantastical: Creating Immersive, Culturally Relevant Learning Experiences with Augmented Reality and Large Language Models       & 2 & 3      \\

61         &   CHI'24          & EmoEden: Applying Generative Artificial Intelligence to Emotional Learning for Children with High-Function Autism        & 2 & 2        \\

62       &   CHI'24          & How AI Processing Delays Foster Creativity: Exploring Research Question Co-Creation with an LLM-based Agent      & 2 & 3     \\

63          &   CHI'24          & Neural Canvas: Supporting Scenic Design Prototyping by Integrating 3D Sketching and Generative AI         & 1 & 3       \\

64          &   CHI'24          & Beyond the Waiting Room: Patient's Perspectives on the Conversational Nuances of Pre-Consultation Chatbots      & 2 & 2       \\

65          &   CHI'24          & Ivie: Lightweight Anchored Explanations of Just-Generated Code        & 2 & 3      \\

66         &   DIS'24          & The CoExplorer Technology Probe: A Generative AI-Powered Adaptive Interface to Support Intentionality in Planning and Running Video Meetings       & 2 & 2        \\

67         &   DIS'24          & DesignPrompt: Using Multimodal Interaction for Design Exploration with Generative AI          & 1 & 3      \\

68         &   DIS'24          & DanceGen: Supporting Choreography Ideation and Prototyping with Generative AI     & 2 & 3           \\

69          &   DIS'24          & PromptInfuser: How Tightly Coupling AI and UI Design Impacts Designers’ Workflows         & 2 & 3    \\

70        &   DIS'24          & GlassMail: Towards Personalised Wearable Assistant for On-the-Go Email Creation on Smart Glasses        & 1 & 2       \\

71         &   DIS'24          & Not Just Novelty: A Longitudinal Study on Utility and Customization of an AI Workflow  & 1 &2           \\

72          &   DIS'24          & Co-Creating Question-and-Answer Style Articles with Large Language Models for Research Promotion     & 2 & 3      \\

73          &   CHI'23          & Understanding the Benefits and Challenges of Deploying Conversational AI Leveraging Large Language Models for Public Health Intervention     & 2 & 2       \\

74          &   CHI'23          & Co-Writing Screenplays and Theatre Scripts with Language Models: Evaluation By Industry Professionals     & 1 & 3         \\

75          &   CHI'23           & Co-Writing with Opinionated Language Models Affects Users' Views     & 2 & 2      \\

76         &   CHI'23          & "What It Wants Me To Say": Bridging the Abstraction Gap Between End-User Programmers and Code-Generating Large Language Models  & 2 & 3        \\

77        &   CHI'23          & AngleKindling: Supporting Journalistic Angle Ideation with Large Language Models     & 2 & 3        \\

78         &   CHI'23         & PopBlends: Strategies for Conceptual Blending with Large Language Models        & 2 & 3    \\

79          &   DIS'23         & Metaphorian: Leveraging Large Language Models to Support Extended Metaphor Creation for Science Writing        & 2 & 2     \\

80          &   DIS'23         & 3DALL-E: Integrating Text-to-Image AI in 3D Design Workflows        & 1 & 3     \\

81        &   CHI'22         & TaleBrush: Sketching Stories with Generative Pretrained Language Models       & 1 & 3      \\

82         &   CHI'22         & Discovering the Syntax and Strategies of Natural Language Programming with Generative Language Models      & 2 & 3       \\

83          &   CHI'22         & Stylette: Styling the Web with Natural Language       & 1 & 3     \\

84         &    CHI'22         & TypeAnywhere: A QWERTY-Based Text Entry Solution for Ubiquitous Computing      & 1 & 2  \\

85          &   DIS'22         & Sparks: Inspiration for Science Writing using Language Models     & 1 & 3     \\

\end{longtable}
\label{tab: artifactlist}
\newpage
\section{294 individual capabilities}


\begin{longtable}{p{0.45\textwidth} p{0.45\textwidth}}
\caption{294 capability list}
\label{tab:capabilitylist} \\

\toprule
\textbf{294 Capabilities} & \textbf{33 Capability Clusters} \\
\midrule
\endfirsthead

\toprule
\textbf{294 Capabilities} & \textbf{33 Capability Clusters} \\
\midrule
\endhead

\midrule
\multicolumn{2}{r}{{Continued on next page}} \\
\bottomrule
\endfoot

\bottomrule
\endlastfoot
summary: output summary of a product description                                                             & summarize into a few sentences based on documents, stories, or dialogue                           \\
correlate: output similar clusters of a topic similarity                                                     & find similar documents based on a conversation, a group of documents, or an item description      \\
Answer: output popular search criteria based on a product                                                    & answer a question about a product, based on a product description                                 \\
criteria: output an explanation of the product based on the criterias                                        & write a description based on keywords                                                             \\
interpret: output the criterias from most relevant to the topic                                              & find similar documents based on a conversation, a group of documents, or an item description      \\
interpret: output the criterias from frequently considered criterias                                         & find similar keywords or document based on dialog, documents, or description (16)                 \\
Rank: output most proper criteria among criterias                                                            & Rank keyword suggestions based on a point in prompt (2)                                           \\
Identify: output paper elements (problems, methods, findings) from a scientific paper                        & Identify sections or elements (problems, methods) based on research paper                         \\
summary: output summary of collection of abstracts                                                           & summarize into a few sentences based on documents, stories, or dialogue                           \\
correlate: output a group of similar papers from a single paper + corpus                                     & find similar documents based on a conversation, a group of documents, or an item description      \\
correlate: output similarities from a group of papers                                                        &                                                                                                   \\
paraphrase: output a paragraph description from a list of similarities (overlapping keywords)                & write a description based on keywords                                                             \\
find similar documents or keywords based on a conversation, a group of documents, or an item description     & Identify sections or elements (problems, methods) based on research paper                         \\
Rank: output top 5 papers from smilar papers                                                                 & find similar documents or keywords based on dialogue, documents, or description                   \\
interpret: output description from a list of similarities (overlapping information)                          & find similar documents based on a conversation, a group of documents, or an item description      \\
Synthesize: output keywords to prompt from user's description of image and user-provided image examples      & summarize into keywords or bullet list based on a document, dialogue, or diary                    \\
correlate: output overlapping featured images from overwhealmed image data                                   & Identify if a person is in an image from a tagged images                                          \\
identify: output speific images from user's description on expecting photo                                   & Identify if a person is in an image from a tagged images                                          \\
summarize: output conversation's shorter version (keywords) from conversation data                           & summarize into keywords or bullet list based on a document, dialogue, or diary                    \\
Filter: output some keywords from abstracted conversatino data                                               & summarize into keywords or bullet list based on a document, dialogue, or diary                    \\
transcribe: output transcript from speech conversation                                                       & transcribe into text from speech                                                                  \\
summarize: output shorter/organized conversation from transcripts                                            & summarize into keywords or bullet list based on a document, dialogue, or diary                    \\
Identify: output similar information with shorter conversation from stored transcript                        & find similar documents or keywords based on dialogue, documents, or description                   \\
vocalize: output audio from stored transcript                                                                & vocalize into speech based on a transcript                                                        \\
program: output JSON format data from text based information                                                 & code into computer code based on a task description                                               \\
Identify: output overlapping contents with target document from relevant documents                           & find similar documents or keywords based on dialogue, documents, or description                   \\
Filter: output relevant information from user's request                                                      & find similar documents or keywords based on dialogue, documents, or description                   \\
Answer: output the answer from user's questions                                                              & answer how to do something based on a question about programming or scientific knowledge          \\
program: output description of persona from JSON                                                             & translate into a description from computational code                                              \\
render: outputs images of personas based on persona description                                              & render a persona image based on a persona description                                             \\
Write: output dialogue from conversation                                                                     & interpret a question to ask someone based on dialog or story                                      \\
context: : output character's construction from the dialogue with user                                       & summarize into keywords or bullet list based on a document, dialogue, or diary                    \\
render: output character's background image from charcter's features(attributes, backstory, context)         & render a persona image based on a persona description                                             \\
Cluster: output categories of dimensions(mood) from user's poem descriptions                                 & summarize into keywords or bullet list based on a document, dialogue, or diary                    \\
Answer: output answers from user's poem description                                                          & answer a question informed by context, based on what was mentioned earlier                        \\
paraphrase: output various tone of answers from one answer                                                   & refine docment's tone of voice based on document and tone request                                 \\
Answer: output responses from specific value of dimension                                                    & answer a question informed by context, based on what was mentioned earlier                        \\
transcribe: output text based transcript from visitor's speech                                               & transcribe into text from speech                                                                  \\
transcribe: output text transcript from user's speech requirements                                           & transcribe into text from speech                                                                  \\
Answer: output audio from user's speech request                                                              & answer a question informed by context, based on what was mentioned earlier                        \\
Filter: ouput text based key points from user's vague requirement transcript                                 & refine into a more explicit and effective prompt based on a vague prompt                          \\
Answer: output the answer from visitor's inquiries                                                           & answer a question informed by context, based on what was mentioned earlier                        \\
context: ouput keywords from supplementary speech guide                                                      & summarize into keywords or bullet list based on a document, dialogue, speech transcript, or diary \\
sentiment understand: output user's interest\&preference from user's audio reaction and rating               & Identify the sentiment from a document, image, or dialog                                          \\
Detect: output user's interest\&preference from user's audio reaction and rating                             & summarize into keywords or bullet list based on a document, dialogue, or diary                    \\
Identify: output specific artwork from user's repeated speech mentions                                       & render an image based on a topic, mood, tone, keywords, or description                            \\
describe: output captions from image segments                                                                & write a description based on image                                                                \\
criteria: output criteria's captions from the image captions                                                 & refine docment's tone of voice based on document and tone request                                 \\
Rank: output top 10 segments from image similarity                                                           & find similar element in the image based on a group of images                                      \\
render: output keywords of image mood from uploaded image                                                    & write a description based on image                                                                \\
refine: output relevant keyword options from selected keywords                                               & refine docment's tone of voice based on document and tone request                                 \\
extend: output longer version of image descriptions from selected keywords                                   & write a description based on keywords                                                             \\
criteria: output image explanations from selected keyword                                                    & render an image based on a topic, mood, tone, keywords, or description                            \\
Identify: output image elements from image explainations                                                     & render an image based on a topic, mood, tone, keywords, or description                            \\
Render: output image variations from image's layout boxes                                                    & render an image that communicates a tone or mood based on an image                                \\
transcribe: output transcript from elcturer's audio                                                          & transcribe into text from speech                                                                  \\
Identify: output transcript keywords from transcript                                                         & summarize into keywords or bullet list based on a document, dialogue, or diary                    \\
Cluster: output categories(learner's understanding level) from transcript keywords                           & Identify sections or elements (problems, methods) based on research paper or dialog               \\
Identify: output criterias from learner's understanding level                                                & Identify sections or elements (problems, methods) based on research paper or dialog               \\
Write: output dialogues from criterias for grading                                                           & write a character's response, based on dialog                                                     \\
summarize: output title of a paper(all the text cotents)                                                     & summarize into a few sentences based on documents, stories, or dialogue                           \\
summarize: output one-line description of paper                                                              & summarize into a few sentences based on documents, stories, or dialogue                           \\
render: output 4 images in different mood from computational prompt                                          & render an image based on a topic, mood, tone, keywords, or description                            \\
summarize: output two-sentence description from paper abstract                                               & summarize into keywords or bullet list based on a document, dialogue, or diary                    \\
Identify: output image elements from input image                                                             & find similar element in the image based on a group of images                                      \\
context: output relationship of image elements from input image                                              & find similar element in the image based on a group of images                                      \\
synthesize: output image descriptions from generated image                                                   & write a description based on image                                                                \\
render: output variations of images from user's preference(like or save)                                        & render an image based on a topic, mood, tone, keywords, or description                            \\
program: output domain-oriented prompts from text-based human prompt                                          & code into computer code based on a task description                                               \\
criteria: output color criterias(warm, cool, netural) from user's description on expecting room mood         & summarize into keywords or bullet list based on a document, dialogue, or diary                    \\
interpret: output color descriptions from 3-color schemes                                                    & write a description based on image                                                                \\
identify: selected color schemes from the previous color schemes                                             & Rank color scheme based on description                                                            \\
compare: output most proper visualization data from overwhealmed random data                                 & Identify sections or elements (problems, methods) based on research paper or dialog               \\
Filter: output selected data from overwhealmed data                                                          & find similar keywords or document based on dialog, documents, or description (16)                 \\
Identify: output UI widgets from user's text prompt                                                          & render an image based on a topic, mood, tone, keywords, or description                            \\
summarize: output shorter keywords from user's expecting image description                                   & summarize into keywords or bullet list based on a document, dialogue, or diary                    \\
Render: output image's plant elements from bounding boxes                                                    & render an image that communicates a tone or mood based on an image                                \\
render: output scene layout from natural language prompts                                                    & render an image based on a topic, mood, tone, keywords, or description                            \\
code: output programming code from user's request on generating plots                                        & code into computer code based on a task description                                               \\
paraphrase: output structured prompt from unstructured user's prompt                                         & refine into a more explicit and effective prompt based on a vague prompt                          \\
summarize: output keywords from interview transcript                                                         & summarize into keywords or bullet list based on a document, dialogue, or diary                    \\
Rank: output most relevant thematic codes from keywords                                                      & interpret explanation based on professional terms                                                 \\
filter: output code-themes from multiple suggestions                                                         & interpret explanation based on professional terms                                                 \\
context: output context-specific questions from contextual audio and text information                        & interpret a question to ask someone based on dialog or story                                      \\
render: output travel blog from images and texts of user's captured moments                                  & write a description based on image                                                                \\
context: output next questions from previous conversation                                                    & interpret a question to ask someone based on dialog or story                                      \\
summarize: output shorter version of document from input travel document                                     & summarize into a few sentences based on documents, stories, or dialogue                           \\
correlate: output relevant painting themes from information similarity with user's image description         & find similar keywords or document based on dialog, documents, or description (16)                 \\
context: output painting themes from user's drawing request with context                                     & summarize into keywords or bullet list based on a document, dialogue, or diary                    \\
render: output image with selected painting theme from original image                                        & render an image that communicates a tone or mood based on an image                                \\
program: output image generation model from painting themes                                                  & code into computer code based on a task description                                               \\
cluster: output group of message types from whole dialogue                                                   & summarize into keywords or bullet list based on a document, dialogue, or diary                    \\
contextual: output counterfactuals for conflict resolution score from conversation with user                 & interpret reason based on a professional knowledge                                                \\
identify: output conflicts from conflict confrontation response                                              & Identify the argument from a document, image, or dialog                                           \\
Write: output simulated responsed from conflict resolution score                                             & answer a question informed by context, based on what was mentioned earlier                        \\
Detect: output irrelavant dialogues from topic similarity on the dialogue                                    & find similar keywords or document based on dialog, documents, or description (16)                 \\
vocalize: output speech from text-response                                                                   & vocalize into speech based on a transcript                                                        \\
cluster: output relevant topic from the dialogues                                                            & find similar keywords or document based on dialog, documents, or description (16)                 \\
Write: output dialogues from previous questions and responses                                                & write a character's response, based on dialog                                                     \\
rank: output five AI agents from personas and perspectives                                                   & Rank personas based on description                                                                \\
Answer: output response options for agents from previous answers                                             & answer a question informed by context, based on what was mentioned earlier                        \\
paraphrase: output persuasive tone prompt from user's general text prompts                                   & refine docment's tone of voice based on document and tone request                                 \\
paraphrase: output tailored questions from user's responses                                                  & interpret a question to ask someone based on dialog or story                                      \\
Write: output analogy from pre-defined text templates                                                        & write a description based on keywords                                                             \\
paraphrase: output news report toned analogy from original analogy                                           & refine docment's tone of voice based on document and tone request                                 \\
Render: output analgy image element options from chosen analogy strategies                                   & render an image based on a topic, mood, tone, keywords, or description                            \\
paraphrase: output numerical values from analogy objects                                                     & refine docment's tone of voice based on document and tone request                                 \\
render: output the related illustration from analogy keywords                                                & render an image based on a topic, mood, tone, keywords, or description                            \\
paraphrase: output proper text prompt from user's vague image description                                    & refine into a more explicit and effective prompt based on a vague prompt                          \\
render: output image from modified text prompt                                                               & render an image based on a topic, mood, tone, keywords, or description                            \\
Rank: output top ranked modifiers from text-to-image prompts                                                 & Rank keyword suggestions based on a point in prompt                                               \\
render: output synonom from children's repeated speech                                                       & refine docment's tone of voice based on document and tone request                                 \\
paraphrase: output antonym from synonom words                                                                & refine docment's tone of voice based on document and tone request                                 \\
context: output definition from a word                                                                       & interpret explanation based on professional terms                                                 \\
cluster: output part-of-speech categories from word definitions                                              & summarize into keywords or bullet list based on a document, dialogue, or diary                    \\
context: output image explanation from input image                                                           & write a description based on image                                                                \\
render: output image from user's text prompt                                                                 & render an image based on a topic, mood, tone, keywords, or description                            \\
Render: output style change of image from original image                                                     & render an image that communicates a tone or mood based on an image                                \\
Identify: output segments from images                                                                        & render an image that communicates a tone or mood based on an image                                \\
Identify: output image layers from image depth information                                                   & render an image that communicates a tone or mood based on an image                                \\
filter: output specific code function from programming code                                                  & answer how to do something based on a question about programming or scientific knowledge          \\
context: output code explanations from programming code                                                      & interpret explanation based on professional terms                                                 \\
context: output code annotation from code explanation                                                        & summarize into keywords or bullet list based on a document, dialogue, or diary                    \\
context: output story related questions from the story                                                       & interpret a question to ask someone based on dialog or story                                      \\
Write: output creative stories based on children's response                                                  & write a story based on a topic or description                                                     \\
context: output math term explanations from the story using math terms                                       & interpret explanation based on professional terms                                                 \\
Answer: output responses from the phase of conversation with patients                                        & answer a question informed by context, based on what was mentioned earlier                        \\
summarize: output shorter dialogue from long dialogue                                                        & summarize into a few sentences based on documents, stories, or dialogue                           \\
Rank: output recommendation on the questions to ask to patients from previous shorter dialogue               & interpret a question to ask someone based on dialog or story                                      \\
paraphrase: output refined sentence from user's sentence tone request                                        & refine docment's tone of voice based on document and tone request                                 \\
summarize: output shorter version of sentence from user's draft                                              & summarize into a few sentences based on documents, stories, or dialogue                           \\
summarize: output summarized labels from the instructions                                                    & summarize into keywords or bullet list based on a document, dialogue, or diary                    \\
paraphrase: output refined sentence from user's pre-determined keywords                                      & write a description based on keywords                                                             \\
context: output next sentence from existing diary sentences and keywords                                     & write a story based on a topic or description                                                     \\
paraphrase: output suggestion of similar sentences from original target sentence                             & refine docment's tone of voice based on document and tone request                                 \\
summarize: output shorter version of diary from paragraphed diary                                            & summarize into a few sentences based on documents, stories, or dialogue                           \\
paraphrase: output revised version of sentence from original sentence and user's prompt explaining intention & refine docment's tone of voice based on document and tone request                                 \\
paraphrase: output grammar revised version of sentence from original sentence                                & refine fix grammar error based on text                                                            \\
summarize: output shorter version of sentence from user's draft                                              & summarize into a few sentences based on documents, stories, or dialogue                           \\
answer: output response for the patients from history feed related to the patients                           & answer a question informed by context, based on what was mentioned earlier                        \\
Answer: output the proper answers from guided topics                                                         & answer how to do something based on a question about programming or scientific knowledge          \\
Write: output dialogues from human-written dialogues with patients                                           & write a character's response, based on dialog                                                     \\
context: output code descriptions in summary from student's code                                             & answer how to do something based on a question about programming or scientific knowledge          \\
Answer: output answer the user's programming questions                                                       & answer how to do something based on a question about programming or scientific knowledge          \\
paraphrase: output revised version of code from user's codes                                                 & refine into a more explicit and effective prompt based on a vague prompt                          \\
Answer: output rapport from conversation with children                                                       & answer a question informed by context, based on what was mentioned earlier                        \\
Identify: output key event labels from conversation with children                                            & summarize into keywords or bullet list based on a document, dialogue, or diary                    \\
summarize: output structured bullet summary from dialogue with children                                      & summarize into keywords or bullet list based on a document, dialogue, or diary                    \\
sentiment understanding: output children's emotion from structured summary                                   & Identify the sentiment from a document, image, or dialog                                          \\
cluster: output negative or positive from conversation with children                                         & Identify the sentiment from a document, image, or dialog                                          \\
Filter: output inapproporate contents from conversation with children                                        & identify an inappropriate or offensive response based on dialogue                                 \\
paraphrase: output positive contents from negative contents in conversation                                  & refine docment's tone of voice based on document and tone request                                 \\
context: output feedback from children's answer (rather correct or wrong)                                    & identify an inappropriate or offensive response based on dialogue                                 \\
Sentiment understanding: output children's emotion from children's previous dialogue                         & Identify the sentiment from a document, image, or dialog                                          \\
Answer: output answer for children's previous question                                                       & answer a question informed by context, based on what was mentioned earlier                        \\
descriptive: output the code explaination from user's code                                                   & translate into a description from computational code                                              \\
Identify: output fragments from similarity of code evaluation                                                & find similar keywords or document based on dialog, documents, or description (16)                 \\
Compare: output the code from user's code and chatbot's code                                                 & Rank needed function for a programmer based on computer code                                      \\
summarize: output summarization of prompt from user's text prompt                                            & summarize into a few sentences based on documents, stories, or dialogue                           \\
summarize: output shorter version description from revising and confirming information                       & summarize into a few sentences based on documents, stories, or dialogue                           \\
paraphrase: output structured tree visualization text from random unorganized information                    & summarize into keywords or bullet list based on a document, dialogue, or diary                    \\
cluster: output group of categories from news                                                                & Identify sections or elements (problems, methods) based on research paper or dialog               \\
context: output explanation and clarification from the propaganda                                            & interpret explanation based on professional terms                                                 \\
Identify: output conversation pattern from the latest dialogues                                              & Identify the sentiment from a document, image, or dialog                                          \\
code: output relevant code from conversational context                                                       & code into computer code based on a task description                                               \\
Write: output dialogues from previous dialogue                                                               & write a character's response, based on dialog                                                     \\
Identify: output drawing's specific elements from children's drawings                                        & find similar element in the image based on a group of images                                      \\
Filter: output characters and scenes from children's image descriptions                                      & summarize into keywords or bullet list based on a document, dialogue, or diary                    \\
code: output system prompt for image generation model from children's image descriptions                     & code into computer code based on a task description                                               \\
transcribe: output text from children's conversation in voice                                                & transcribe into text from speech                                                                  \\
program: output block prompting combiation from questions                                                    & refine into a more explicit and effective prompt based on a vague prompt                          \\
context: output specific knowledge from conversation with children                                           & answer how to do something based on a question about programming or scientific knowledge          \\
program: output structured programming code from block prompting                                             & summarize into keywords or bullet list based on a document, dialogue, or diary                    \\
Compare: output student's score from students' responses                                                     & identify an inappropriate or offensive response based on dialogue                                 \\
Answer: output student's answer from student's questions                                                     & answer how to do something based on a question about programming or scientific knowledge          \\
summarize: output question summarization from user's question                                                & summarize into a few sentences based on documents, stories, or dialogue                           \\
paraphrase: output personalized story letter from gamification mission letters                               & refine docment's tone of voice based on document and tone request                                 \\
Answer: output response from user's question                                                                 & answer how to do something based on a question about programming or scientific knowledge          \\
write: output follow-up questions from previous conversation                                                 & interpret a question to ask someone based on dialog or story                                      \\
answer: output response from conversation with patients                                                      & answer a question informed by context, based on what was mentioned earlier                        \\
Detect: output vague response from conversation with patients                                                & refine into a more explicit and effective prompt based on a vague prompt                          \\
Answer: output emotional answers from patient's previous conversation                                        & answer a question informed by context, based on what was mentioned earlier                        \\
correlate: output related questions to ask from UI components                                                & interpret a question to ask someone based on dialog or story                                      \\
Write: output responses from actionable operation scripts                                                    & answer a question informed by context, based on what was mentioned earlier                        \\
paraphrase: output alternative versions from main story                                                      & refine docment's tone of voice based on document and tone request                                 \\
paraphrase: output alternative versions of story from children's target works                                & refine docment's tone of voice based on document and tone request                                 \\
summarize: output 3-sentence abstracts from existing story for chidren                                       & summarize into a few sentences based on documents, stories, or dialogue                           \\
summarize: output sentence blocks(keywords) from 3-sentences                                                 & summarize into keywords or bullet list based on a document, dialogue, or diary                    \\
correlate: output the specific relevant argument from the description                                        & Identify the argument from a document, image, or dialog                                           \\
context: output keywords to explain from a sentence                                                          & summarize into keywords or bullet list based on a document, dialogue, or diary                    \\
paraphrase: output refined version segments from words                                                       & refine docment's tone of voice based on document and tone request                                 \\
program: output action code function from input refined segments                                             & code into computer code based on a task description                                               \\
Rank: output proper know-action from action code functions                                                   & Rank needed function for a programmer based on computer code                                      \\
Answer: output answers from user's questions                                                                 & answer a question informed by context, based on what was mentioned earlier                        \\
context: output specific claims from sentences                                                               & Identify the argument from a document, image, or dialog                                           \\
context: output questions from specific claims                                                               & interpret a question to ask someone based on dialog or story                                      \\
answer: output questions based on the request                                                                & interpret a question to ask someone based on dialog or story                                      \\
sentiment understanding: output semantic context from pre-assembled database                                 & Identify the sentiment from a document, image, or dialog                                          \\
identify: output keywords for the search from the prompt                                                     & summarize into keywords or bullet list based on a document, dialogue, or diary                    \\
identify: output the name of the app                                                                         & answer a question about a product, based on a product description                                 \\
decode: output natural language test instructions from actionable steps                                      & translate into a description from computational code                                              \\
context: output personal questions from conversation with user                                               & interpret a question to ask someone based on dialog or story                                      \\
context: output practical questions from conversation with user                                              & interpret a question to ask someone based on dialog or story                                      \\
identify: output metaphor keywords from users' NL prompt                                                     & summarize into keywords or bullet list based on a document, dialogue, or diary                    \\
write: output metaphorical text from storyline visualization                                                 & write a description based on image                                                                \\
render: output 512*512 resolution images from user's NL prompt                                               & render an image based on a topic, mood, tone, keywords, or description                            \\
context: output text depiction from generated images                                                         & write a description based on image                                                                \\
program: output codes from intermediary variable                                                             & code into computer code based on a task description                                               \\
program: output code function from user's natural language                                                   & code into computer code based on a task description                                               \\
decode: output textual prompts from code function                                                            & translate into a description from computational code                                              \\
cluster: output type of goal from rhetorical problem or writing goal                                         & interpret reason based on a professional knowledge                                                \\
Write: output phase from rhetorical problem or writing goal                                                  & write a description based on keywords                                                             \\
paraphrase: output revised version from user's draft                                                         & refine docment's tone of voice based on document and tone request                                 \\
summarize: output shorter sentence from user's draft                                                         & summarize into a few sentences based on documents, stories, or dialogue                           \\
interpret: output student's comment options from previous selected options                                   & answer a question informed by context, based on what was mentioned earlier                        \\
Write: output sentences from student's comment options                                                       & write a character's response, based on dialog                                                     \\
context: output hint-text from GUI prompt                                                                    & summarize into keywords or bullet list based on a document, dialogue, or diary                    \\
summarize: output hint-text(extracted keyword) from suggested input contents                                 & summarize into keywords or bullet list based on a document, dialogue, or diary                    \\
context: output hint-text from the feedback                                                                  & summarize into keywords or bullet list based on a document, dialogue, or diary                    \\
paraphrase: output diverse version of research ideation sentences from idea                                  & refine docment's tone of voice based on document and tone request                                 \\
correlate: output relevant sentence from the research sentence                                               & find similar keywords or document based on dialog, documents, or description (16)                 \\
summarize: output summary of related wrok from the relevant research papers                                  & summarize into a few sentences based on documents, stories, or dialogue                           \\
summary: output the shorter version of document from research paper                                          & summarize into a few sentences based on documents, stories, or dialogue                           \\
summarize: output shorter version of document from research paper                                            & summarize into a few sentences based on documents, stories, or dialogue                           \\
Compare: output relevant sentences from existing works                                                       & find similar keywords or document based on dialog, documents, or description (16)                 \\
paraphrase: output clean transcript from original transcript                                                 & refine fix grammar error based on text                                                            \\
summarize: output summarization from transcript                                                              & summarize into a few sentences based on documents, stories, or dialogue                           \\
summarize: output keywords from summary                                                                      & summarize into keywords or bullet list based on a document, dialogue, or diary                    \\
summarize: output sentences from transcription                                                               & summarize into a few sentences based on documents, stories, or dialogue                           \\
paraphrase: output synonym from the keywords                                                                 & refine docment's tone of voice based on document and tone request                                 \\
paraphrase: output related antonym words from synonym words                                                  & refine docment's tone of voice based on document and tone request                                 \\
render: output dream related image from explaination of user's dream                                         & render an image based on a topic, mood, tone, keywords, or description                            \\
Rank: output most proper brainstorming design concepts from keywords                                         & Rank keyword suggestions based on a point in prompt                                               \\
render: output image from text prompt and edge map                                                           & render an image based on a topic, mood, tone, keywords, or description                            \\
synthesize: output music description from the uploaded image                                                 & write a description based on image                                                                \\
render: output the image from the textual prompts                                                            & render an image based on a topic, mood, tone, keywords, or description                            \\
answer: output the response from textual prompts                                                             & answer a question informed by context, based on what was mentioned earlier                        \\
paraphrase: output the constructive feedback from prompts                                                    & refine into a more explicit and effective prompt based on a vague prompt                          \\
context: output the evaluation result from the violation                                                     & refine docment's tone of voice based on document and tone request                                 \\
Answer: output give context related answer from the user's questions and previous answers                    & answer a question informed by context, based on what was mentioned earlier                        \\
context: output scenario based dialogues from selected scenario and persona                                  & write a character's response, based on dialog                                                     \\
Write: output creative edge scenarios from simulation options                                                & write a story based on a topic or description                                                     \\

Interpret: output questions from research & Interpret explanation based on professional terms  \\
Answer: output given context related answer from generated questions
  & Answer how-to based on question about programming or science \\ Interpret: output follow-up questions from previous questions and answers

   &  Interpret a question to ask someone based on dialog or story \\Transcribe: output text from speech

    & Transcribe into text from speech \\Identify: output sentiment key points  from email

     & Identify sentiment from document, image, or dialog  \\Summarize: output keywords from user's request

  & Summarize into keywords/bullets based on document, dialog, or diary \\Write: output email from keywords

   &Write description based on keywords  \\Transcribe: output text from speech

    & Transcribe into text from speech  \\Render: output 3D dance image from text description
  & Render image based on topic, mood, tone, keywords, or description \\Render: output 2D thumbnail images from 3D dance images

    & Render image that communicates tone or mood based on image \\
 Render: output 3D dance image from different sequenced images &  Render image that communicates tone or mood based on image\\Render: output 3D dance images from stopped motion images

    & Render image that communicates tone or mood based on image \\Render: output design inspired images from various modality prompts
  & Render image based on topic, mood, tone, keywords, or description \\Render: output design inspired images from text descriptions

    & Render image based on topic, mood, tone, keywords, or description \\Render: output image from machine-tranlated prompts

    & Render image based on topic, mood, tone, keywords, or description  \\Summarize: output a sentence meeting goal from meeting description
  & Summarize into few sentences based on documents, stories, or dialog \\ Summarize: output meeting summaries from meeting calender information

    & Summarize into few sentences based on documents, stories, or dialog \\ Identify: output meeting topics from emails
  & Identify sections/elements (problems, methods) based on research paper or dialog \\Identify: output key-point of feedback from user's answers

    &  Identify sections/elements (problems, methods) based on research paper or dialog \\Summarize: output corresponding information from user's UI design

        & Summarize into few sentences based on documents, stories, or dialog \\Rank: output relevant recommendation from previous works
  & Rank keyword suggestions based on a point in prompt \\Refine: output multiple answers in different tone of voice from prompts

    &  Refine document's tone of voice  based on document and tone\\ Refine: output multiple toned versions from original article
  & Refine into more explicit/effective prompt based on vague prompt \\Write: output hook from original article

    & Write description based on keywords \\ Write: output anecdotes from original article

        & Write description based on keywords \\Summarize: output main points from the article's sections
  &Summarize into few sentences based on documents, stories, or dialog   \\Find similar: output multiple controversies from given press

    &  Find similar keywords or document based on dialog, documents, or description \\ Identify: output key-points from press
  & Identify sentiment from document, image, or dialog  \\Find similar: output negative key-points from press

    & Find similar keywords or document based on dialog, documents, or description  \\Find similar: output similar posts from discussion
& Find similar keywords or document based on dialog, documents, or description  \\Answer: output reponse from dialog history

    & Answer question based on earlier mentions  \\Write: output script from user's input
& Write story based on topic or description  \\Find similar: output similar suggestion from theatre scripts

    & Find similar keywords or document based on dialog, documents, or description \\Write: output code from natural language prompt

    & Write code based on task description  \\Render: output product scenes from natural language prompt

    & Render image based on topic, mood, tone, keywords, or description \\Identify: output user's sentiment from selected keywords

    & Identify sentiment from document, image, or dialog \\Write: output text description from generated images

    &Write description based on image   \\ Find similar: output similar concepts from existing images and design

    & Find similar: output similar concepts from existing images and design \\ Identify: output connecting concepts from properties

    & Identify sentiment from document, image, or dialog \\Refine: output synonoms from different themes

    &Refine document's tone of voice  based on document and tone   \\Find similar: output similar scientific words from user's scientific words

    & Find similar keywords or document based on dialog, documents, or description  \\Find similar: output relevant subscientific words from metaphor word

    &  Find similar keywords or document based on dialog, documents, or description \\Find similar: output alternative words from previous words

    & Find similar keywords or document based on dialog, documents, or description  \\Summarize: output design keywords from user's natural language prompt

    & Summarize into keywords/bullets based on document, dialog, or diary  \\Render: output 3D components from user's 2D image prompt

    & Render image that communicates tone or mood based on image \\Render: output similar 4 images from text prompt
    
    & Render image based on topic, mood, tone, keywords, or description  \\Render: output similar 4 images from 2D image prompts

    & Render image that communicates tone or mood based on image \\Refine: output different tone of voice prompt from guide prompts
    
    & Refine into more explicit/effective prompt based on vague prompt  \\Write: output complete sentence from sentence with blank

    & Write story based on topic or description \\Write: output creative story from user generated plots 

        & Write story based on topic or description  \\Identify: output sentence in similar level of fortuen from whole text

    & Identify sections/elements (problems, methods) based on research paper or dialog  \\Transcribe: output text from speech

    &  Transcribe into text from speech \\Write: output CSS properties from user's natural language prompt

    & Write code based on task description  \\Summarize: output key CSS components from generated CSS codes

        & Summarize into keywords/bullets based on document, dialog, or diary \\ Code: output HTML code from natural language

    &  Write code based on task description

    \\Code: output JavaScript code from natural language

    & Write code based on task description  
    
    \\
    
    Summarize: output summarized keywords from documents

    & Summarize into keywords and bullets based on document, dialog, or diary  
    
    \\
    
    Write: output text or several characters from probability

        &  Write story based on topic or description \\

\bottomrule
\end{longtable}
\label{tab: 241capabilitytable}
\twocolumn

\end{document}